\documentclass[useAMS,usenatbib]{mn2e}
\usepackage{graphicx}
\usepackage{subfigure}
\usepackage{amsmath}
\usepackage{amssymb}
\usepackage{cancel}
\usepackage{hyperref}

\title[The microphysics of D-type expansion]{On the relative importance of different microphysics on the D-type expansion of galactic H\,II regions}
\author[T. J. Haworth et al.]
{\parbox{\textwidth}{T. J. Haworth$^{1}$\thanks{E-mail: \texttt{thaworth@ast.cam.ac.uk}}, T. J. Harries$^{2}$, D. M. Acreman$^{2}$ and T. G. Bisbas$^{3,4}$ 
}\vspace{0.4cm}\\
\parbox{\textwidth}{$^{1}$Institute of Astronomy, Madingley Road, Cambridge, CB3 0HA, UK\\
$^{2}$School of Physics, University of Exeter, Stocker Road, Exeter, EX4 4QL, UK\\
$^{3}$Department of Physics and Astronomy, University College London. 132 Hamstead Road, Kings Cross, London, NW1 3EE, UK\\
$^{4}$Max Planck Institut f{\"u}r Extraterrestrische Physik, Giessenbachstrasse 1, 85748 Garching, Germany\\
}}
\begin{document}

\date{Accepted ???. Received ???; in original form ???}

\pagerange{\pageref{firstpage}--\pageref{lastpage}} \pubyear{2014}

\maketitle

\label{firstpage}

\begin{abstract}
Radiation hydrodynamics (RHD) simulations are used to study many astrophysical phenomena, however they require the use of simplified radiation transport and thermal prescriptions to reduce computational cost. In this paper we present a systematic study of the importance of microphysical processes in RHD simulations using the example of D-type H\,\textsc{ii} region expansion. We compare the simplest hydrogen--only models with those that include: ionisation of H, He, C, N, O, S and Ne, different gas metallicity, non-LTE metal line blanketed stellar spectral models of varying metallicity, radiation pressure, dust and treatment of photodissociation regions. Each of these processes are explicitly treated using modern numerical methods rather than parameterisation. In line with expectations, changes due to microphysics in either the effective number of ionising photons or the thermal structure of the gas lead to differences in D--type expansion. In general we find that more realistic calculations lead to the onset of D--type expansion at smaller radii and a slower subsequent expansion. Simulations of star forming regions using simplified microphysics  are therefore likely overestimating the strength of radiative feedback. We find that both variations in gas metallicity and the inclusion of dust can affect the ionisation front evolution at the 10--20 per cent level over 500\,kyr, which could substantially modify the results of simplified 3D models including feedback. Stellar metallicity, radiation pressure and the inclusion of photodissociation regions are all less significant effects at the 1 per cent level or less, rendering them of minor importance in the modelling the dynamical evolution of H\,\textsc{ii} regions. 
\end{abstract}


\section{Introduction}
\label{introduction}
Radiative feedback has the potential to influence the morphological evolution of star forming regions and to induce or inhibit star formation \citep[for a recent review see][]{dalerev}. The radiation field from massive stars ionizes the surrounding gas, heating it and causing expansion of the hot high pressure region  \citep{1978ppim.book.....S, 2006ApJ...646..240H, 2012MNRAS.419L..39R}. This can result in the dispersal of material, hindering  the formation of stars, or can collect/destabilize material and potentially trigger star formation. A large number of factors contribute to the effectiveness of radiative feedback (making models difficult, because there is a huge parameter space of initial conditions and microphysical complexity) and feedback processes are expected to take place over timescales of order 1\,Myr in systems of complex geometry (making observations difficult, due to projection effects and the challenge of constructing a dynamical picture from a single snapshot). Consequently the impact of radiative feedback on star formation is still not clear, particularly in any quantitative sense \citep[see e.g.][]{2015MNRAS.450.1199D}.

Radiation hydrodynamics (RHD) calculations have been used to investigate the effect of radiative feedback in a host of ``star formation'' scenarios. The external irradiation of an isolated cloud \citep[e.g.][]{2007MNRAS.377..383E, 2009MNRAS.393...21G, 2009MNRAS.392..964R, 2011ApJ...736..142B, 2011BSRSL..80..391M, 2012A&A...538A..31T,  2012MNRAS.422.1352D, 2012MNRAS.420..562H, 2014MNRAS.444.1221K}, collect and collapse \citep[e.g.][]{2007MNRAS.375.1291D} and the radiatively driven evolution of turbulent media \citep[e.g.][]{2009ApJ...694L..26G, 2011MNRAS.414.1747A, 2012MNRAS.427..625W, 2012A&A...546A..33T}  have been studied using numerical models, providing a phenomenological picture of the impact of radiative feedback on the evolution of both the gaseous and stellar content of star forming regions. 

These RHD models all necessarily use (to varying extents) a simplified treatment of radiation transport/photoionisation due to the complexity of the microphysics and finite computational resources available. In recent years some effort has been invested into understanding how important these assumptions are. For example \cite{2012MNRAS.420..562H} investigated the effect of including polychromatic and
diffuse field radiation in models of the radiatively driven implosion of clouds, finding that the diffuse radiation field can significantly
modify the results. \cite{2012A&A...546A..33T} also found that the assumption of photoionisation equilibrium can affect the results of RHD calculations of the external irradiation of a turbulent medium. They found that non-equilibrium photoionisation was required to detatch the tips of elephant trunks to form Bok globules \citep{1947ApJ...105..255B}. More recently \cite{2014MNRAS.439.2990S} investigated the effect of radiation pressure, finding that it is a secondary effect compared to photoionisation. \cite{2015MNRAS.448.3248G} also studied the relative impacts of winds, ionising radiation and supernova feedback in a series of 1D models, which they used to summarise the energetic feedback into the ISM from a 15\,M$_{\odot}$ star.

There are further approximations that have not been formally investigated, for example the assumption that the gas is hydrogen--only (thus neglecting cooling from forbidden line transitions and heating and cooling from helium) and using a simplified thermal balance that calculates the temperature as a function of hydrogen ionisation fraction \citep[though e.g.][compare the heating and cooling rates]{1999RMxAA..35..123R, 2006ApJ...647..397M, 2010MNRAS.403..714M, 2006MNRAS.369..143M,  2012A&A...546A..33T}. There are also variations in the gas or stellar metallicity, dust and photodissociation regions/FUV heating. The impact of these approximations must be investigated to understand by how much and why simple models differ from more complex ones. 

We can easily illustrate how different approximations might be expected to affect the evolution of H\,\textsc{ii} regions by considering the classic system of a massive star at the centre of a uniform density medium. The star rapidly ionises a sphere of the surrounding gas. If the gas is hydrogen only and we invoke the on-the-spot (OTS) approximation, under which the diffuse radiation field is neglected, then the radius of this initial bubble is the Str\"{o}mgren radius 
\begin{equation}
	r_{\textrm{s}} = \left(\frac{3N_{\textrm{ly}}}{4\pi n_{\textrm{e}}^2 \alpha_{\textrm{B}}}\right)^{1/3}
	\label{stromgren}
\end{equation}
where $N_{\textrm{ly}}$, $n_{\textrm{e}}$ and $\alpha_{\textrm{B}}$ are the number of ionising photons emitted by the star, gas electron density and the case B recombination coefficient (for recombinations into all states other than the ground). The subsequent D-type expansion of the H\,\textsc{ii} region was described by \cite{1978ppim.book.....S} as
\begin{equation}
	r_{\textrm{I}} = r_{\textrm{s}}\left(1 + \frac{7\, c_{\textrm{I}} t}{4\, r_{\textrm{s}}}\right)^{4/7}
	\label{spitzer}
\end{equation}
where $c_{\textrm{I}}$ is the sound speed in the ionised gas. 
If we consider a case B recombination coefficient sensitive to temperature as $T^{-0.8}$ (see equation \ref{alphaB}) then the Str\"{o}mgren radius varies as $T^{0.27}$. This will affect the expansion rate at times when $\frac{7\, c_{\textrm{I}} t}{4\, r_{\textrm{s}}} \lesssim 1$. At times when $\frac{7\, c_{\textrm{I}} t}{4\, r_{\textrm{s}}} >> 1$ we expect that the expansion of an H\,\textsc{ii} region will vary with the temperature in the ionised gas as $r_{\textrm{I}} \propto T_{\textrm{I}}^{2/7}t^{4/7}$ and the number of ionising photons as $r_I\propto N_{ly}^{1/7}t^{4/7}$. Although these dependencies are quite weak, departures from simple estimates in these quantities may add up over time to give substantial H\,\textsc{ii} region expansion differences. Furthermore, if we improve our model by considering gas that is not hydrogen only and includes the diffuse field then both the recombination coefficient and the electron density will change, further modifying the Str\"{o}mgren radius which affects the expansion.  There are hence multiple factors that could affect the expansion of H\,\textsc{ii} regions when improving the microphysical treatment.

In this paper we aim to  explore how different stellar and gas metallicities, radiation pressure, dust and photodissocation regions affect the D--type expansion of H\,\textsc{ii} regions, which we interpret in terms of the simple analytic expectations mentioned above. Specifically we will  investigate the effects of these processes in \textit{galactic} H\,\textsc{ii} regions rather than those studied in the epoch of reionisation and make the distinction based on the gas densities, gas constituents and the ionising fluxes from the sources considered. 

\section{Numerical method}
\label{nummeth}
We use the Monte Carlo radiation transport and hydrodynamics code \textsc{torus} \citep{2000MNRAS.315..722H, 2004MNRAS.351.1134K, 2010MNRAS.407..986R, 2012MNRAS.426..203H} to perform the calculations in this paper. The \textsc{torus} RHD algorithm uses operator splitting to separate grid--based hydrodynamics and Monte Carlo photoionisation \citep{2012MNRAS.420..562H}. The primary advantage of this method is that all of the features available to a dedicated Monte Carlo radiation transport code are available in RHD calculations. The disadvantage is that this approach is computationally expensive, however it can be efficiently parallelized using a range of techniques \citep{2015MNRAS.448.3156H}. Details and testing of the RHD algorithm are provided in \cite{2012MNRAS.420..562H} however we include a summary here for completeness since this paper predominantly explores different physical processes that the code can include. We also use the coupled \textsc{torus-3dpdr} code, which is discussed in detail in \cite{TORUS3DPDRpaper} and also summarised here.

\subsection{Hydrodynamics}
We use a flux conserving, finite difference hydrodynamics algorithm. It is total variation diminishing (TVD) and makes use of the van Leer flux limiter \citep{vanleer} and a Rhie-Chow interpolation scheme to prevent odd--even decoupling \citep{1983AIAAJ..21.1525R}. \textsc{torus} is capable of treating point source \citep{2015MNRAS.448.3156H} and self-gravity, the latter of which is calculated using a multigrid method, though we do not include gravity in the models in this paper.

\subsection{Photoionisation and thermal balance}
\label{photo}
\textsc{torus} uses a photoionisation scheme similar to that of \cite{2003MNRAS.340.1136E} and \cite{2004MNRAS.348.1337W} which in turn are based on the method presented by \cite{1999A&A...344..282L}. Packets of photons at constant frequency and that carry constant energy $\epsilon$ (but whose members vary in number with frequency) are propagated throughout the computational grid. As they traverse grid cells they trace a path length $l$ and modify the time--averaged radiation energy density $U$ in the cell by 
\begin{equation}
	dU = \frac{4\pi J_{\nu}}{c} d \nu = \frac{\epsilon}{c\Delta t}
        \frac{1}{V} \sum_{d\nu} l.
\label{energydensityMC}
\end{equation}
where $V$ is the cell volume, $c$ is the speed of light, $J_{\nu}$ is the mean intensity and $\Delta t$ is the time over which the averaging takes place. The update to the radiation energy density takes place following any photon packet `event' which, as well as absorption, includes traversal of a cell boundary. Following an absorption the photon packet is re-emitted with random frequency and direction under the principle of detailed balance, continuing a random walk through the grid until it escapes. The spectrum for diffuse field photons is constructed using 1000 frequencies that strategically sample Lyman continuum and helium ionising photons,  hydrogen recombination lines, forbidden lines and the dust continuum (the diffuse field is thus dependent on the species included and ionisation and temperature state of the gas). Once all photon packets in the calculation have escaped the radiation energy density is used to solve the ionisation balance equation \citep{1989agna.book.....O} which, in terms of Monte Carlo estimators, is given by
\begin{equation}
	\frac{n(X^{i+1})}{n(X^i)} = \frac{ \epsilon}{\Delta t V \alpha(X^i) n_{\rm{e}}} \sum\frac{l a_{\nu}(X^i)}{  h\nu}
	\label{ionBalanceMC}
\end{equation}
where $n(X^i)$, $\alpha(X^i)$, $a_{\nu}(X^i)$ and $n_{\rm{e}}$ are the electron number density, recombination coefficient and absorption cross section of ion $X^i$ and the electron density respectively. 
The approach of using the crossing of cell boundaries as a photon packet event has the advantage that photon packets contribute to the estimate of the radiation field without having to undergo absorption events, thus even very optically thin regions are properly sampled. 

For the simplest models in this paper we assume that the gas is either entirely atomic or  ionised hydrogen. The radiation field is monochromatic and we use the OTS approximation, including the same case B recombination coefficient as used by \cite{2009A&A...497..649B}
\begin{equation}
	\alpha_B = 2.7\times10^{-13}\left(\frac{T}{10^4}\right)^{-0.8}.
	\label{alphaB}
\end{equation}
Our simple models employ a common \citep[e.g.][]{2009MNRAS.393...21G, 2011ApJ...736..142B} simplified thermal balance calculation where the temperature in cell ${j}$ is a two--temperature interpolation function of the ionisation fraction of atomic hydrogen $\eta_{{j}}$
\begin{equation}
	T_{{j}} = T_{\rm{n}} +\eta_{{j}}(T_{\rm{io}} - T_{\rm{n}})
	\label{quicktherm}
\end{equation}
where $T_{\rm{n}}$ is the prescribed temperature of fully neutral gas and $T_{\rm{io}}$ the prescribed temperature of fully ionized gas (10 and $10^4$\,K respectively). 
We retain the assumption of photoionisation equilibrium in all models. 

In contrast to the simple  calculations, in the detailed photoionisation models we include a range of atomic constituents: hydrogen, helium, carbon, nitrogen, oxygen, neon and sulphur, for which we solve the ionisation balance using equation \ref{ionBalanceMC}. The levels that we treat for metals are C (I--IV), N (I--IV), O (I--III), Ne (II--III), S (II--IV). The hydrogen, helium and C\,IV recombination rates used by \textsc{torus} are calculated based on \cite{1996ApJS..103..467V}. Other radiative recombination rates are calculated using fits to the results of \cite{1983A&A...126...75N}, \cite{1991A&A...251..680P} or \cite{1982ApJS...48...95S}. The photoionisation cross sections of all atomic species in this paper are calculated using the \textsc{phfit2} routine from \cite{1996ApJ...465..487V}. We note that we assume rather than explicitly calculate the abundance of helium and metals and use this assumed abundance to calculate the ionisation structure. 

The detailed photoionisation model temperature is calculated by finding the temperature at which the heating and cooling rates in each cell match. The gas heating rate from hydrogen and helium in a given cell is calculated based on the sum of trajectories of photon packets through the cell  \citep[][]{2004MNRAS.348.1337W}. The dust heating is calculated separately, but using the same method \citep[][]{1999A&A...344..282L}. 

The cooling rate is initially calculated for the maximum and minimum allowed temperatures in the calculation ($3\times10^{4}$\,K and 10\,K respectively by default in \textsc{torus}). This is then refined by bisection until the cooling rate matches the heating rate. For gas, the cooling processes considered are those from free--free radiation, hydrogen and helium recombination and collisional excitation of hydrogen and metals. For dust, there is blackbody radiative cooling. We iterate over the ionisation and thermal balance calculations until the fractional change in both the ionisation fraction and temperature is less than $10^{-2}$. Gas and dust are thermally (but not dynamically) decoupled, having their thermal balance solved independently.

In the detailed calculations that are not investigating the effect of dust we set the dust to gas ratio to the negligibly low value of $10^{-20}$.

\subsection{\textsc{OSTAR2002} Spectral models from \textsc{TLUSTY}}
\label{ostar2002}
\begin{figure}
	\hspace{-20pt}
	\includegraphics[width=9.1cm]{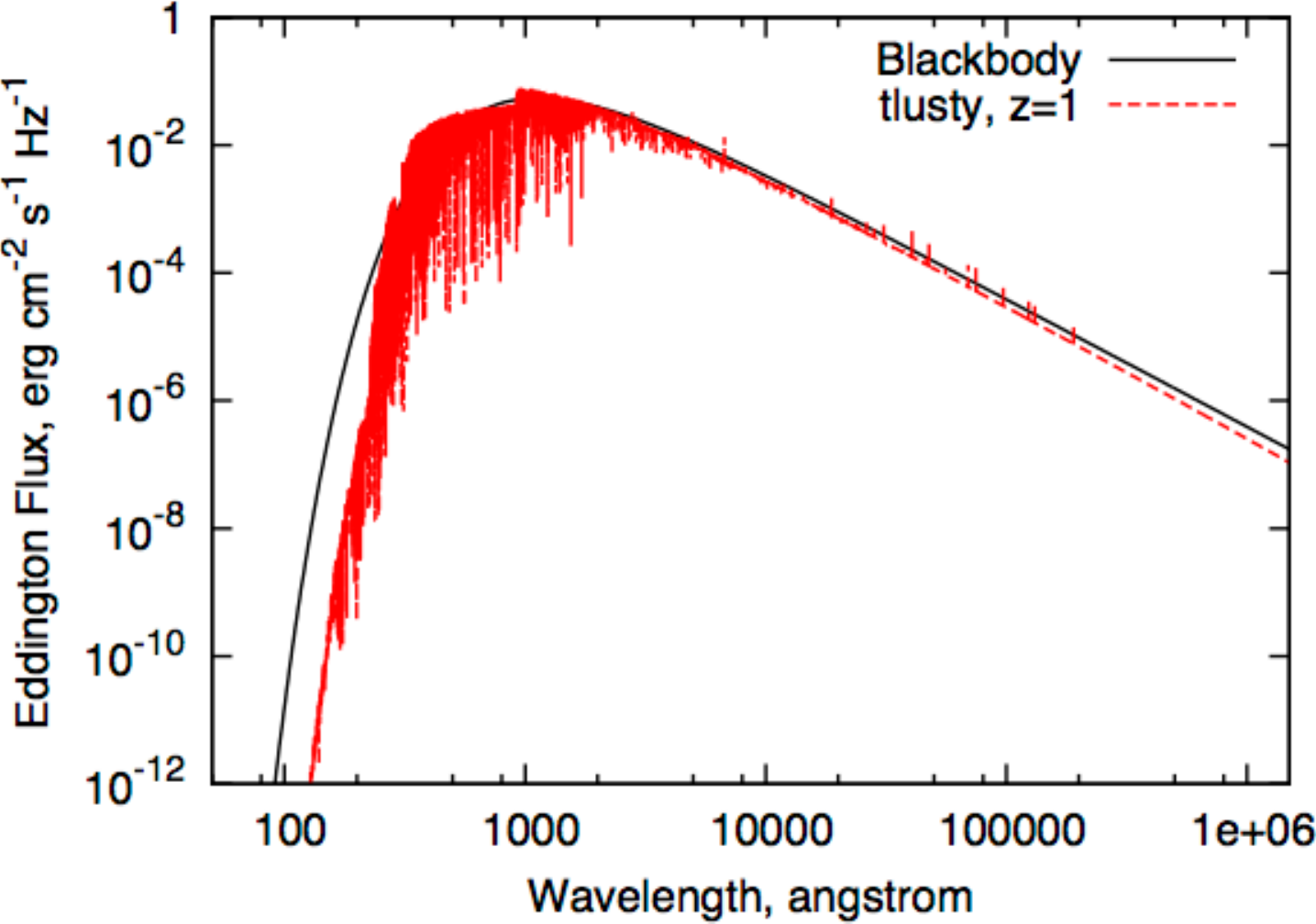}

	\hspace{-20pt}	
	\includegraphics[width=9.1cm]{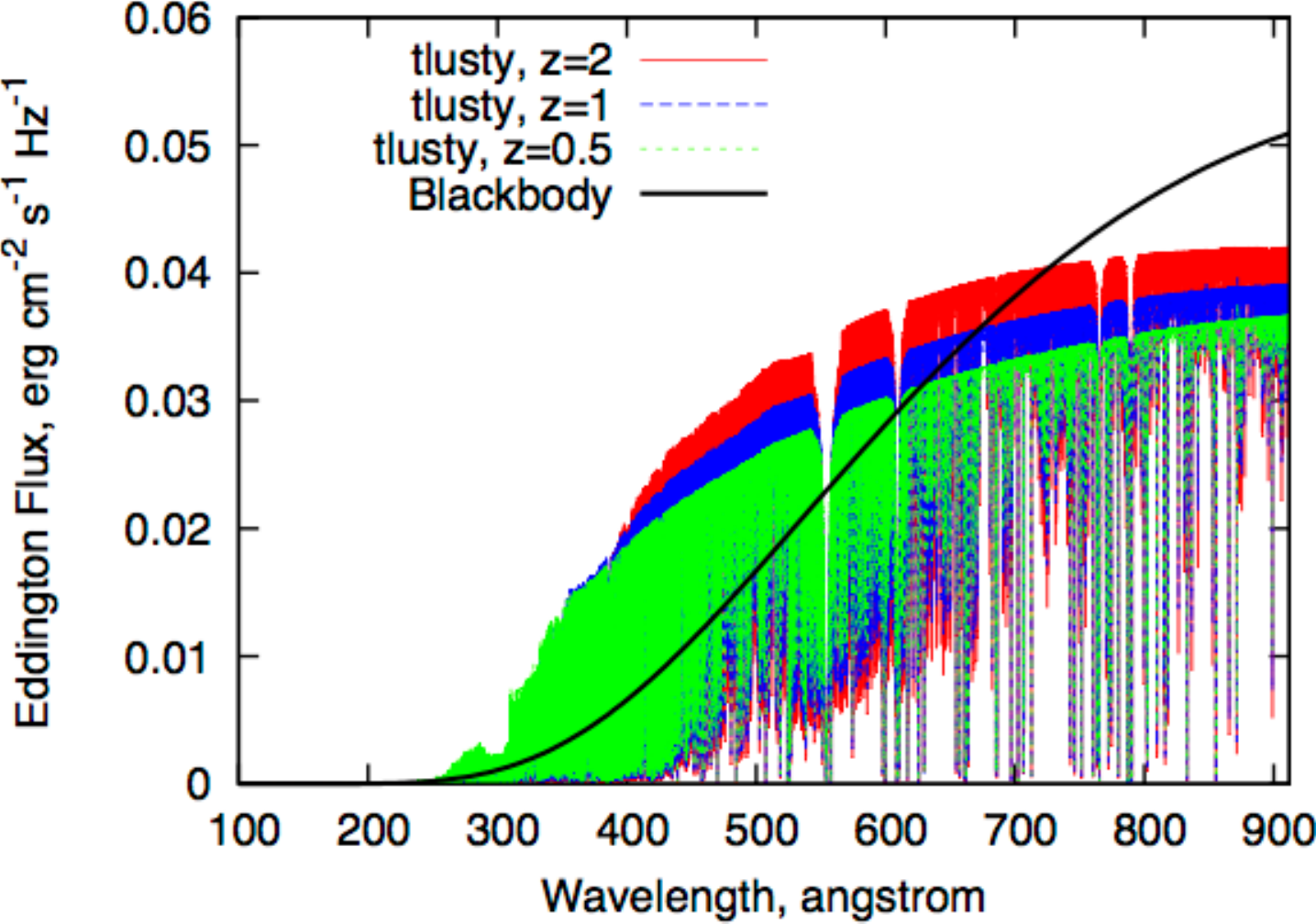}

	\caption{Blackbody and \textsc{tlusty} spectra for a 45000\,K star of radius 10.9\,R$_{\odot}$. The top panel shows a broad range of the spectrum for a blackbody and star of solar metalliciity. The lower panel shows the ionising component of the spectrum for a blackbody and stars of $Z=0.5, 1$ and $2\,Z_\odot$. Note that the flux at the stellar surface is $4\pi$ times the Eddington flux. } 
	\label{BBvsSpec}
\end{figure}

For stellar spectral models, we use the non-LTE, metal line blanketed, plane parallel radiation transport and hydrostatic equilibrium ``OSTAR2002'' models of \cite{2003ApJS..146..417L}, calculated using the code \textsc{tlusty}. We use three sets of grids for stellar metallicities of $Z=0.5, 1$ and 2\,$Z_{\odot}$. Each grid consists of 69 models, spanning temperatures from 27500\,K to 55000\,K and surface gravities from $3.0 \leq \log(g) \leq 4.75$. The actual spectrum used in a calculation is derived by interpolation between the two grid spectra with properties closest to that of the star in our model. In Figure \ref{BBvsSpec} we show an example blackbody spectum and that interpolated from OSTAR2002 for a star at 45000\,K, a radius of 10.9\,R$_{\odot}$ and various stellar metallicities. The upper panel shows the spectrum over a large wavelength range and the lower panel shows the spectrum from the Lyman limit. 

\subsubsection{Frequency sampling of the stellar and diffuse field spectra}
\label{nuSample}
\textsc{torus} converts the source spectrum into a cumulative probability distribution as a function of frequency.  Generating a random number in the range 0:1 and mapping this onto the cumulative probability distribution then yields a frequency with probability appropriate to the source spectrum. The \textsc{tlusty} spectra consist of 193434 frequencies per model and the blackbody spectra consist of 1000 frequencies. The blackbody spectrum is logarithmically sampled from 10 to $10^7$\,\AA. 


\subsection{Radiation pressure}
\cite{2015MNRAS.448.3156H} discusses and tests the treatment of radiation pressure by \textsc{torus} in detail. In summary the radiation pressure force is calculated using Monte Carlo estimators. This estimate is calculated as photon packets are propagated over the grid in the photoionisation component of the calculation and so if already doing the photoionisation, is essentially obtained for no additional computational cost.  This technique accounts for polychromatic radiation and anisotropic scattering and works in both the free-streaming and optically thick regimes. The calculated radiation pressure appears as an additional force term in the hydrodynamic component of the calculation.

\subsection{Dust}
\label{dustdesc}
Historically, \textsc{torus} has been used to study discs around young stars where the gas and dust can be assumed to be thermally coupled. For modelling of H\,\textsc{ii} regions, which are at much lower densities, we must thermally decouple the dust from the gas. This has been done by other codes \citep[e.g.][]{2005MNRAS.362.1038E, 2013ARep...57..573P}, but by \textsc{torus} for the first time in this paper. 

We assume spherical silicate dust particles that follow a standard interstellar medium power law size distribution \citep[e.g.][]{1977ApJ...217..425M} of the form where the number of grains of radius $a$ is

\begin{equation}
	n(a) = c \,a^{-q} \textrm{e}^{-a}
\end{equation}
where $c$ and $q$ are constants.

The dust optical constants are taken from \cite{1984ApJ...285...89D}. We use a pre-tabulated Mie-scattering phase matrix. At present our dust treatment does not include photoelectric heating or resonant line transfer. We assess the impact of this approximation by comparing with the more advanced (in terms of photoionisation) Monte Carlo photoionization code \textsc{mocassin} \citep{2003MNRAS.340.1136E, 2005MNRAS.362.1038E} which does include these dust processes.

The model that we use for testing is the 1D HII40 Lexington benchmark \citep[see][]{1995aelm.conf...83F, 2003MNRAS.340.1136E, 2012MNRAS.420..562H}, only with the inclusion of dust. The dust to gas mass ratio is $10^{-2}$ and we use Draine silicate grains. The  density is 100\,m$_\textrm{H}$\,cm$^{-3}$ and the metal abundances used by both codes are as given in Table \ref{LexingtonParams}. 

\begin{figure}
	\hspace{-20pt}
	\includegraphics[width=9cm]{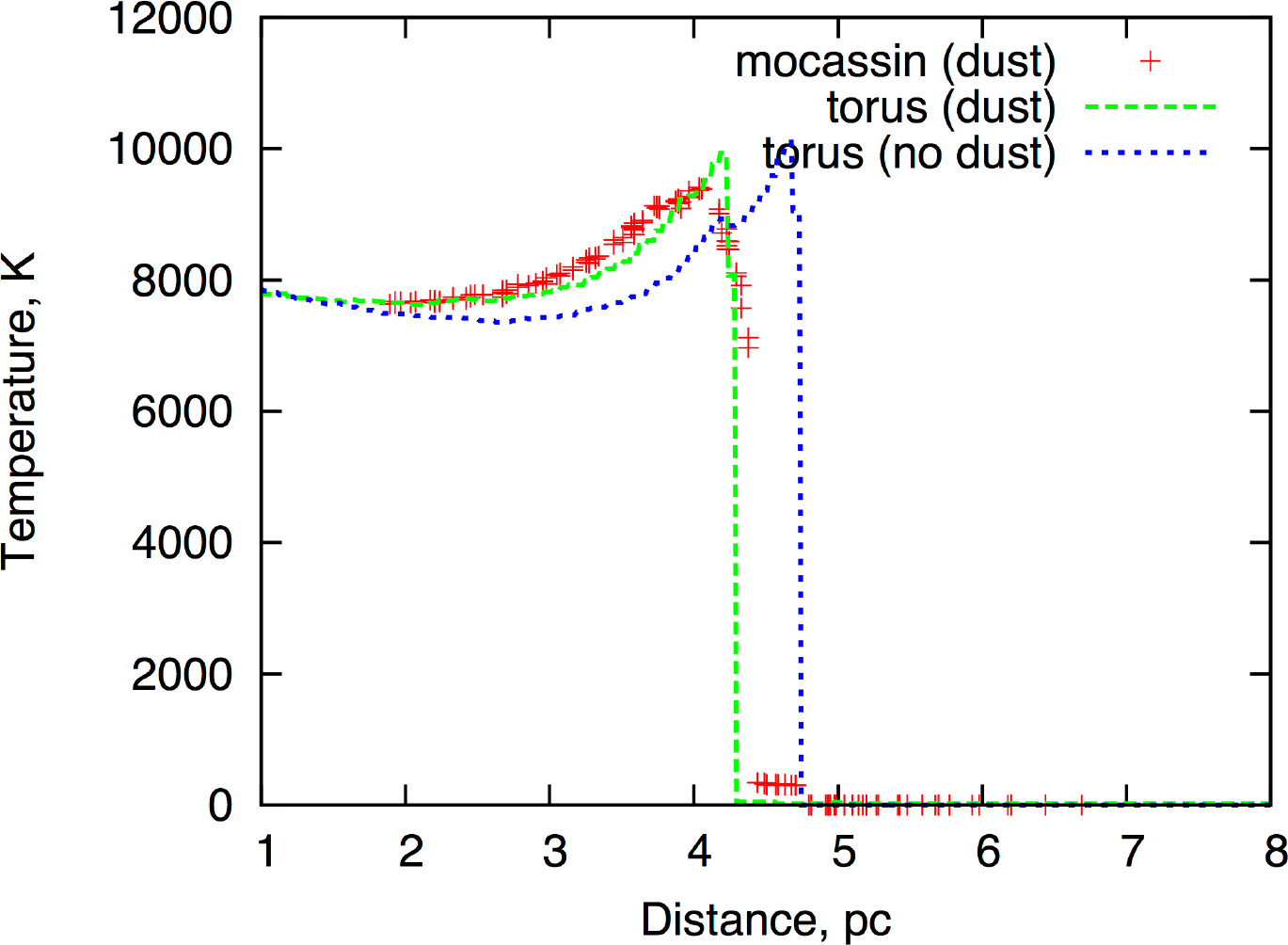}
	\caption{The temperature distribution in a uniform medium about an ionising source calculated by \textsc{torus} both with (green dashed line) and without dust (blue dotted line). Included is also the result including dust computed by  \textsc{mocassin} (red crosses).}
	\label{dusttest}
\end{figure}

A comparison of the gas temperature distribution (all that really matters for these dynamic calculations, as opposed to the line intensities) as computed by \textsc{torus} and \textsc{moccasin} is given in Figure \ref{dusttest}. Clearly the thermally decoupled gas and dust model from \textsc{torus} calculates a very similar H\,\textsc{ii} region radius and temperature to \textsc{mocassin}. The extent of the ionised gas has been reduced compared to the simulation in which there is no dust. Beyond the ionisation front there is slight heating (of order tens of Kelvin) in the neutral gas by \textsc{torus}, but this is weaker than in the \textsc{mocassin} calculation, where there neutral gas is heated to $\sim300$\,K. We do not expect this downstream heating to have much effect on the dynamics compared to the more dramatic effect on the ionisation front radius. We note that this additional heating in the \textsc{mocassin} calculation is very similar to photodissociation region heating, which will be investigated in this paper, so we will be able to gauge its impact.

\subsection{TORUS-3DPDR}
We recently coupled \textsc{torus} with the \textsc{3d-pdr} code of \cite{2012MNRAS.427.2100B}, the details and testing of which are given thoroughly in \cite{TORUS3DPDRpaper}. This approach is novel since the UV field used in the PDR calculation is determined explicitly by the Monte Carlo radiation transport (typically the Draine field is a free parameter for PDR codes). The coupled code is fully integrated, with the PDR calculations taking place on the \textsc{torus} grid. This means that we can calculate the ionisation state of multiple species in the H\,\textsc{ii} region as well as the chemical and thermal structure of the PDR and then use the resulting thermal properties as pressure terms in the hydrodynamics. \cite{2015arXiv150600471M} recently presented a 2D hydrochemical code capable of modelling PDR chemistry and evolving the gas dynamics. The difference with our code is that we can directly calculate the UV field over all space and can also calculate the properties in the H\,\textsc{ii}  region. Unfortunately such a calculation is  computationally very expensive.  

\textsc{torus-3dpdr} uses the most recent UMIST 2012 chemical network database, consisting of 215 species and more than 3000 reactions \citep{2013A&A...550A..36M}. However at present we use a reduced network of 33 species and 330 reactions, primarily to reduce computational cost. Nevertheless, RHD simulations including even this reduced network of species and reactions, in conjunction with photoionisation, are novel. Note that we assume equilibrium in both the H\,\textsc{ii} region and the PDR.

\begin{table}
 \centering
  \caption{H\,\textsc{ii} region expansion model parameters.}
  \label{LexingtonParams}
  \begin{tabular}{@{}l c l@{}}
  \hline
   Variable (Unit) & Value & Description\\
 \hline
   $\textrm{T}_{\rm{eff}} \textrm{(K)}$ & 45000 & Source effective temperature \\
   ${R}_*(R_\odot)$  & 10.9 & Source radius\\
   $M_*(M_\odot)$& 63.8 & Source mass \\
   $\log_{10}(g)$	& 4.17 & Source surface gravity \\
   $\rho$  ($m_H$\,cm$^{-3}$) & 100 & Low density model density\\
   $\rho$ ($m_H$\,cm$^{-3}$) & 500 & High density model density\\
   log$_{10}$(He/H) & $-$1 & Base helium  abundance\\
   log$_{10}$(C/H) & $-$3.66 & Base carbon  abundance ($z_{\rm{g}} =1$)\\
   log$_{10}$(N/H) & $-$4.40 & Base nitrogen abundance ($z_{\rm{g}} =1$)\\
   log$_{10}$(O/H) & $-$3.48 & Base oxygen abundance ($z_{\rm{g}} =1$)\\
   log$_{10}$(Ne/H) & $-$4.30 & Base neon  abundance ($z_{\rm{g}} =1$)\\
   log$_{10}$(S/H) & $-$5.05 & Base sulphur  abundance ($z_{\rm{g}} =1$)\\
   d/g & $1\times10^{-2}$ & Dust to gas mass ratio \\
   a$_{min}$ & 0.005 & Minimum dust grain size \\
   a$_{max}$ & 0.25 & Maximum dust grain size \\
   q & 3.3 & Dust power law index \\   
   L (cm) & 4.4$\times 10^{19}$ & Computational domain size\\
   $n_{\rm{cells}}$ & $256$ & Number of grid cells\\
\hline
\end{tabular}
\end{table}

Modelling of the PDR should give rise to higher temperatures (a few hundred Kelvin rather than 10) at the boundary of the H\,\textsc{ii} region. This will allow us to investigate the effect of heating seen by dust just beyond the ionisation front  by \textsc{mocassin} that is not replicated by \textsc{torus} (see Figure \ref{dusttest}).



\section{Test model specifications}
The model that we use as a testbed is the classic expansion of an H\,\textsc{ii} region about an ionising star in a uniform density medium \citep[][]{1978ppim.book.....S, 1980pim..book.....D, 2006ApJ...646..240H, 2012MNRAS.419L..39R, 2015arXiv150705621B}.  This system is 1D spherically symmetric, reducing computational cost and thus allowing us to incorporate a large range of microphysics in this paper. Specifically we only consider the early phase expansion \citep{2015arXiv150705621B}.

We consider a star of effective temperature 45000\,K, radius 10.9\,$R_\odot$ and mass $63.8$\,M$_\odot$. These stellar parameters are taken from \cite{1998ApJ...501..192D} and is in the regime where non-LTE model atmospheres are required. For a model star with these parameters we consider the expansion of the H\,\textsc{ii} region into two different ambient densities of 100 and 500\,m$_{\textrm{H}}$\,cm$^{-3}$ which we refer to as low and high density respectively. For this scenario we run a host of models with different physical prescriptions: simple photoionisation and full photoionisation  (see section 2.2), non-LTE spectral models, different gas metallicities, radiation pressure, dust and treatment of photodissociation regions. We study the gas and stellar metallicity effects separately so that we can determine which, if either, the dynamics is most sensitive to. In reality the gas and stellar metallicities will not be independent. A summary of the physical parameters in the model is given in Table \ref{LexingtonParams}.

\section{Results and Discussion}
\begin{figure*}
	\includegraphics[width=5cm]{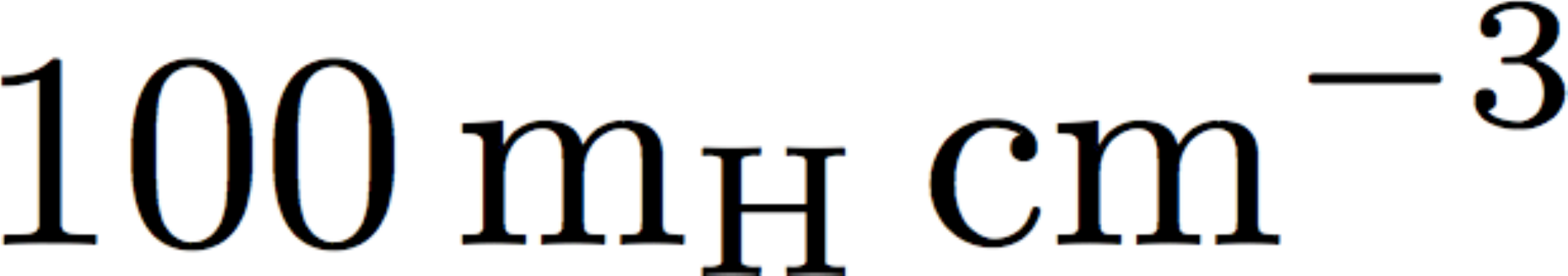}

	\hspace{-15pt}
	\includegraphics[width=9cm]{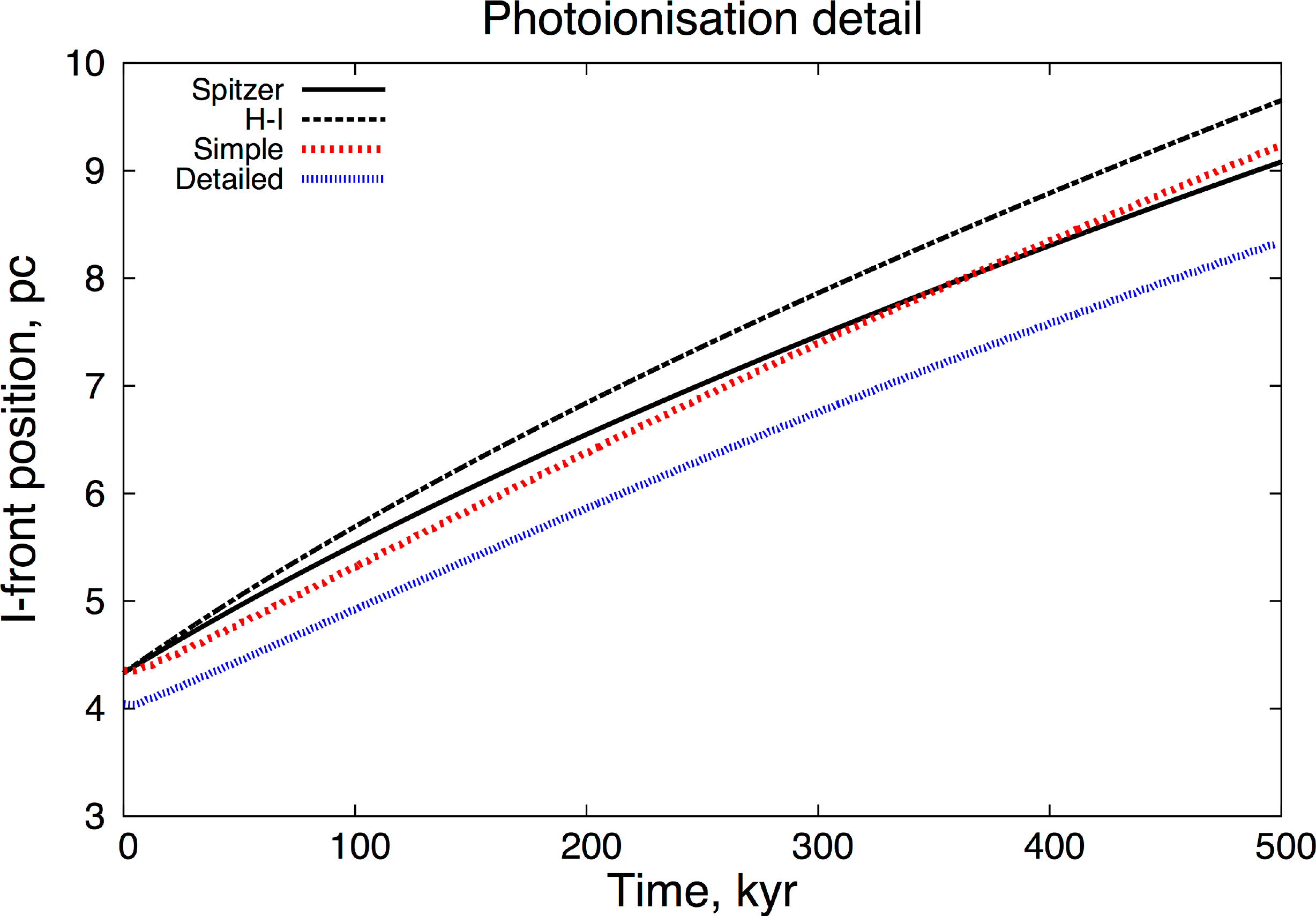}	
	\includegraphics[width=9cm]{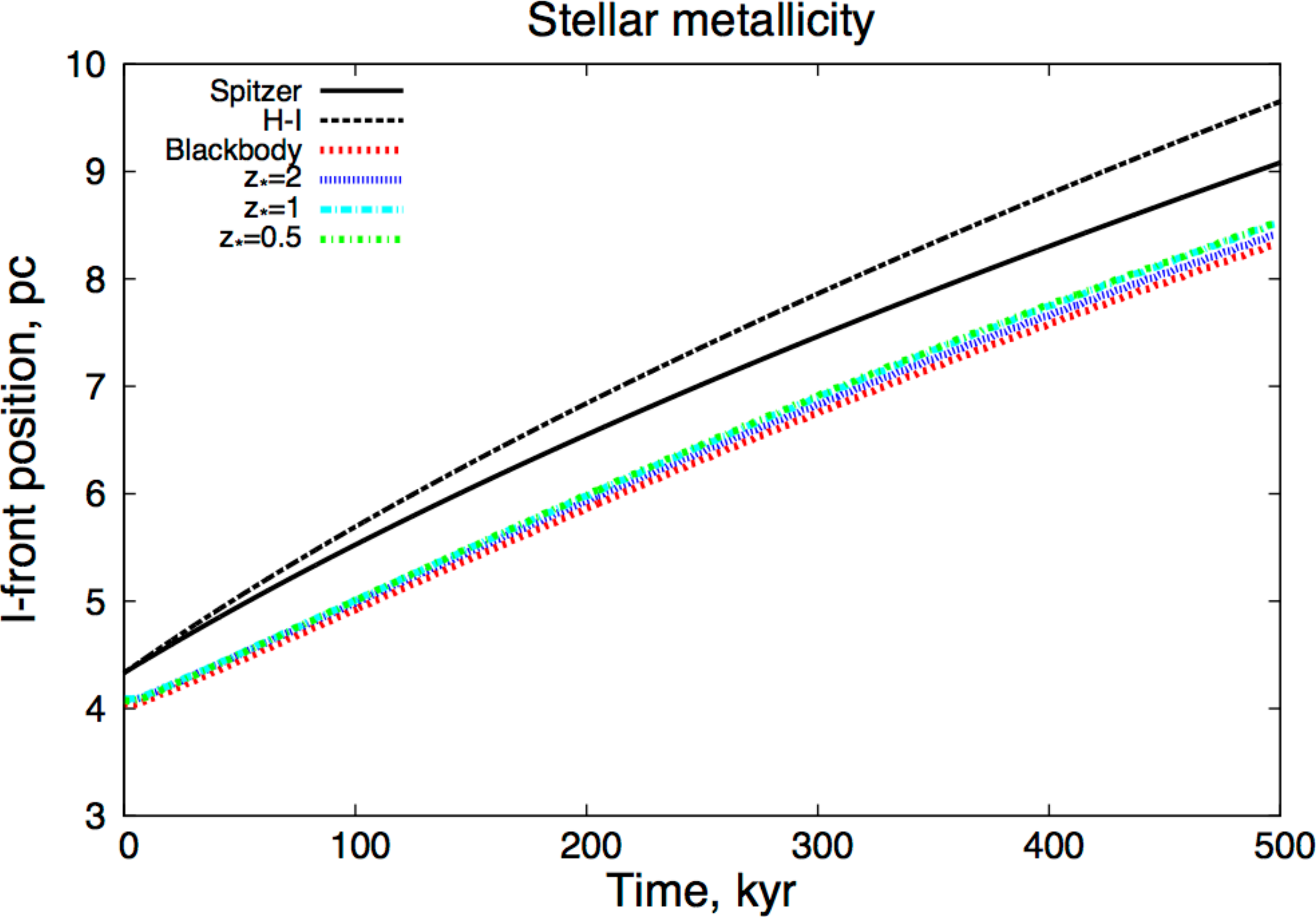}

	\vspace{30pt}
	\hspace{-15pt}	
	\includegraphics[width=9cm]{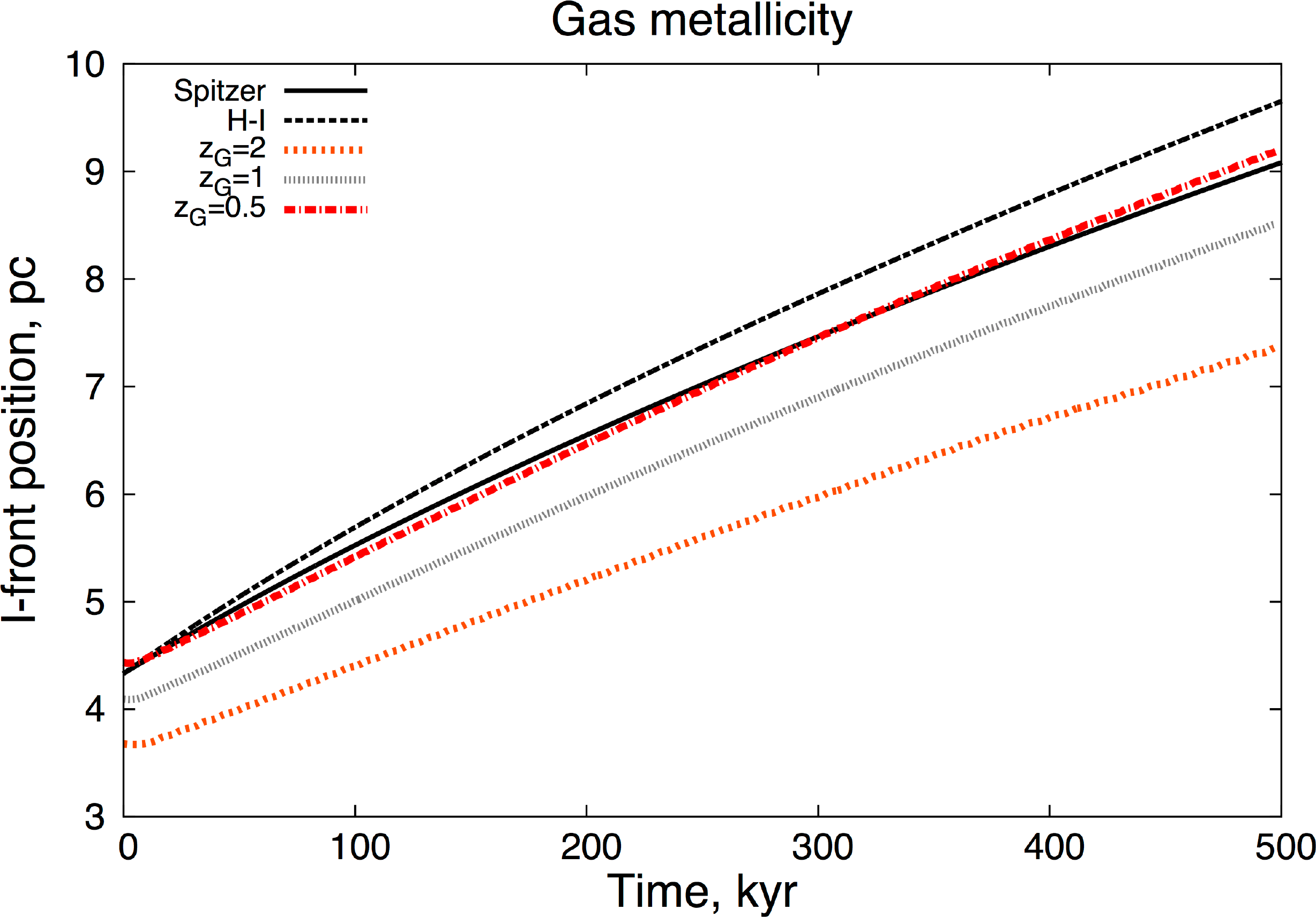}
	\includegraphics[width=9cm]{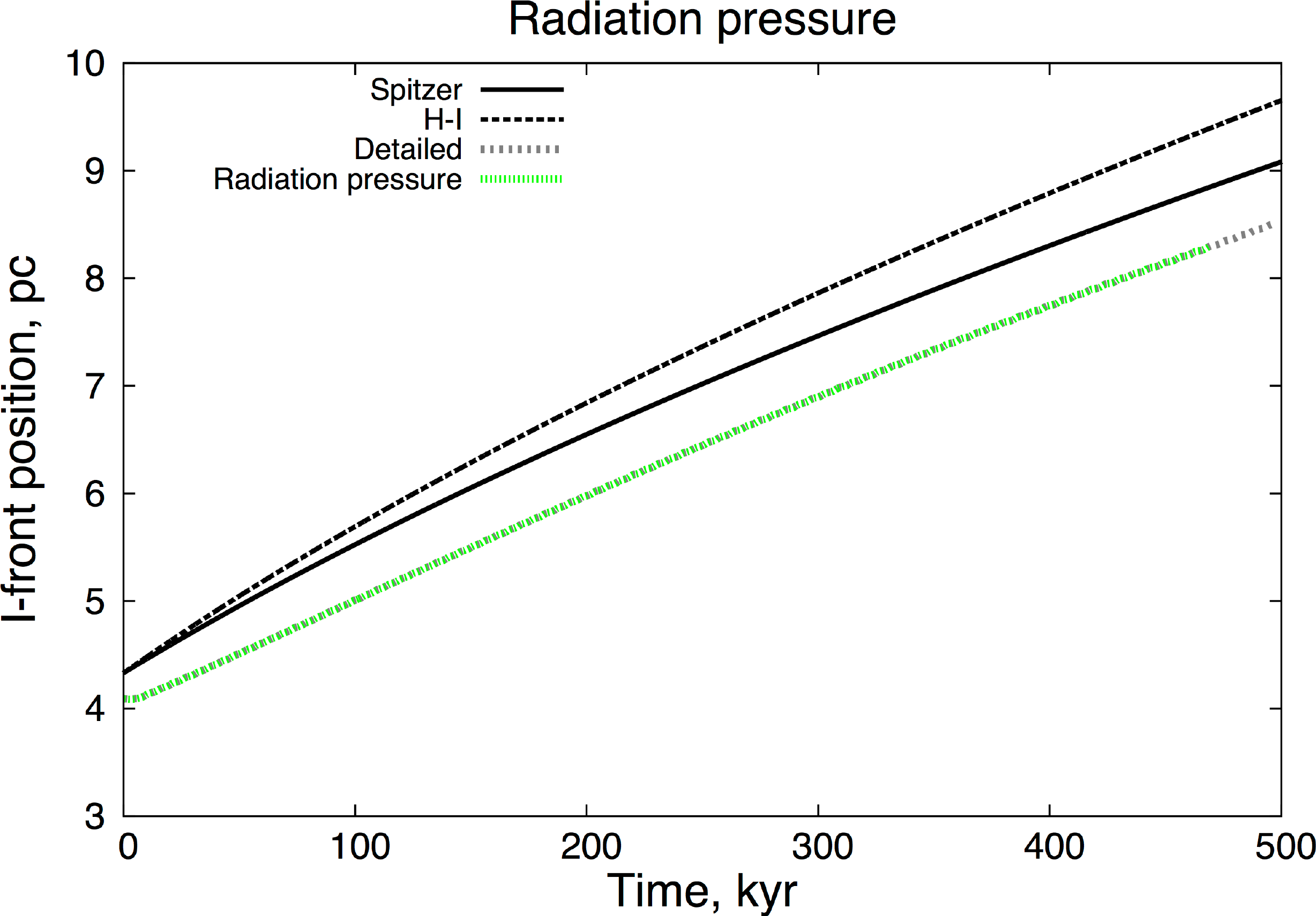}	

	\vspace{30pt}
	\hspace{-15pt}
	\includegraphics[width=9cm]{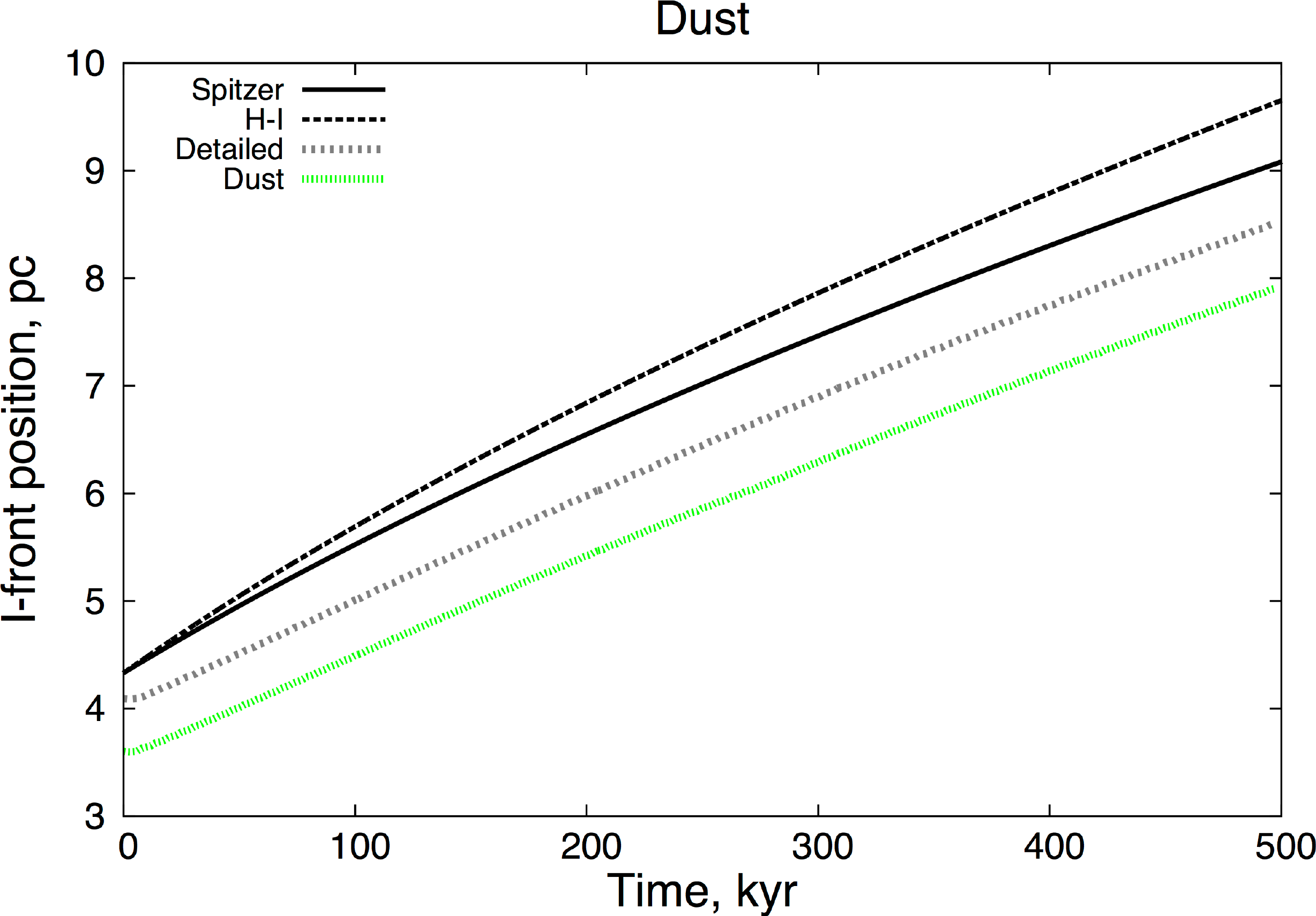}		
	\includegraphics[width=9cm]{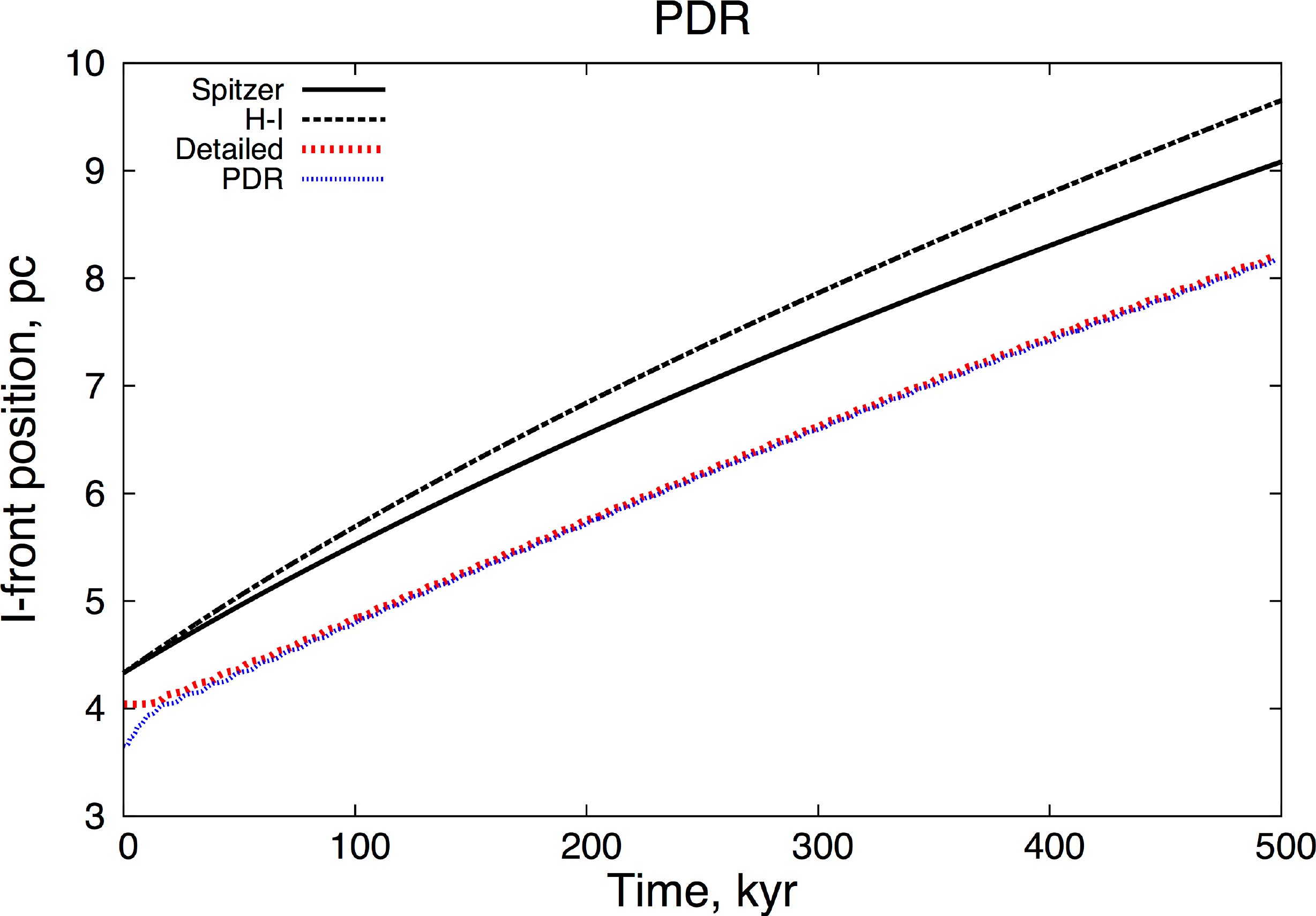}		
	\caption{The ionisation front position as a function of time for all of the 100\,$m_{\textrm{H}}$\,cm$^{-3}$ models in this paper. The top left panel compares the most simplified model with standard Monte Carlo photoionisation plus hydrodynamics. The top right panel includes detailed spectral models. Middle left varies the gas metallicity, middle right includes radiation pressure, bottom left includes dust and the bottom right includes PDR treatment. }
	\label{all}
\end{figure*}

\begin{figure*}
	\includegraphics[width=5cm]{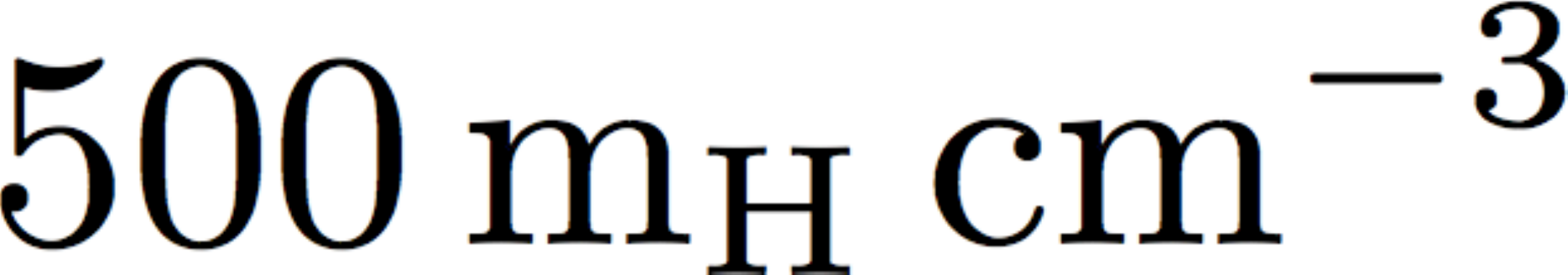}	

	\hspace{-15pt}
	\includegraphics[width=9cm]{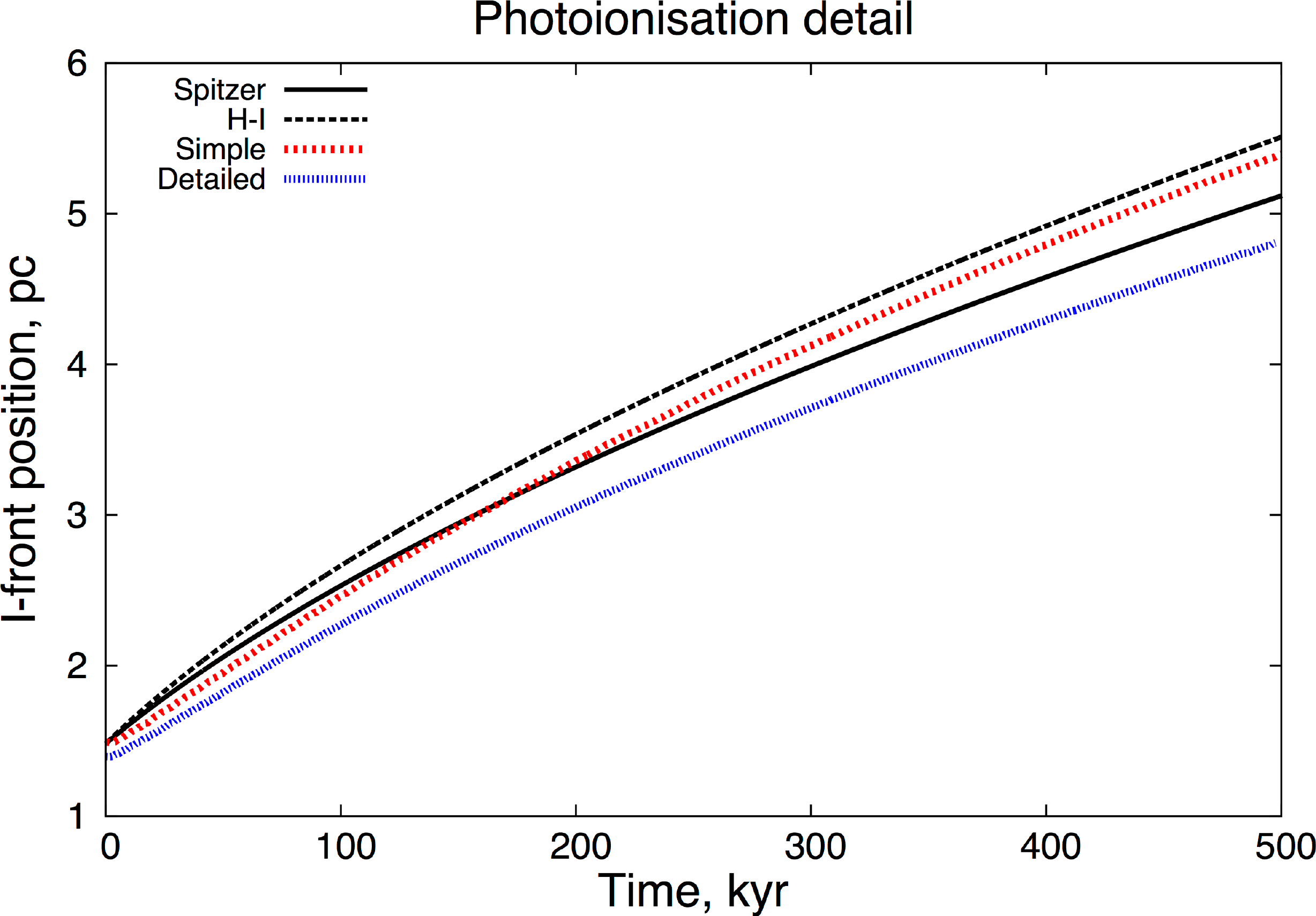}	
	\includegraphics[width=9cm]{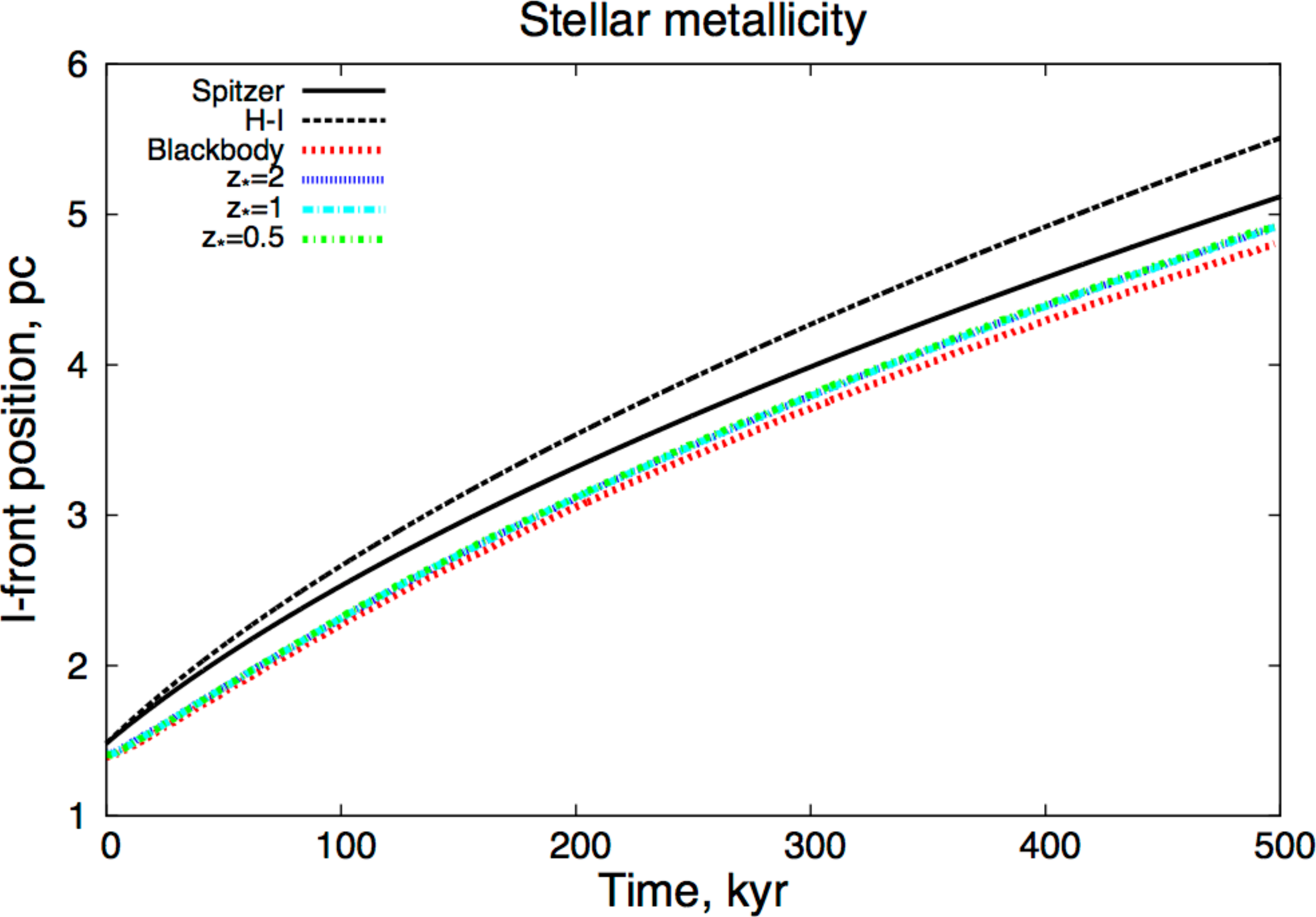}

	\vspace{30pt}
	\hspace{-15pt}	
	\includegraphics[width=9cm]{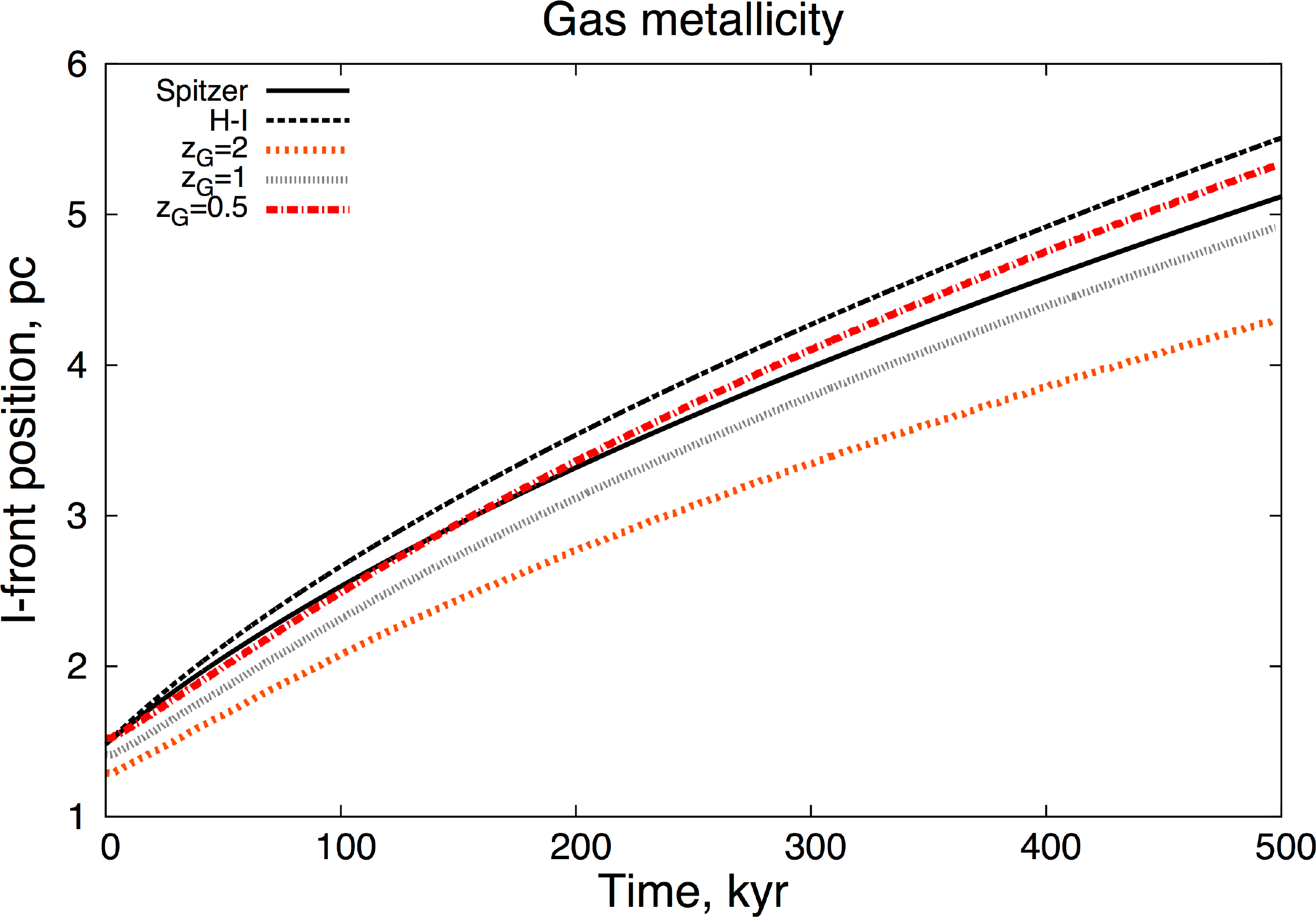}
	\includegraphics[width=9cm]{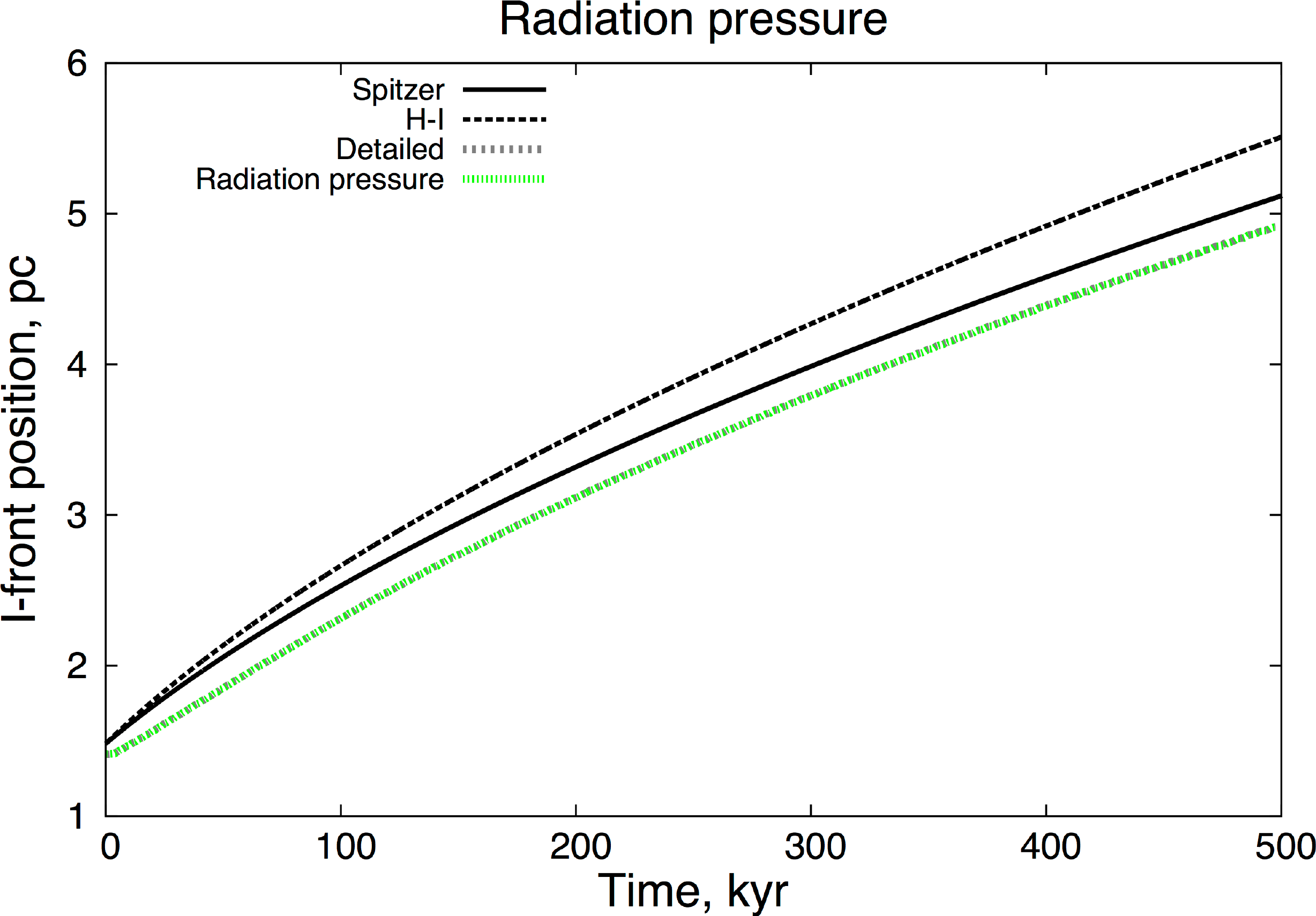}

	\vspace{30pt}
	\hspace{-15pt}	
	\includegraphics[width=9cm]{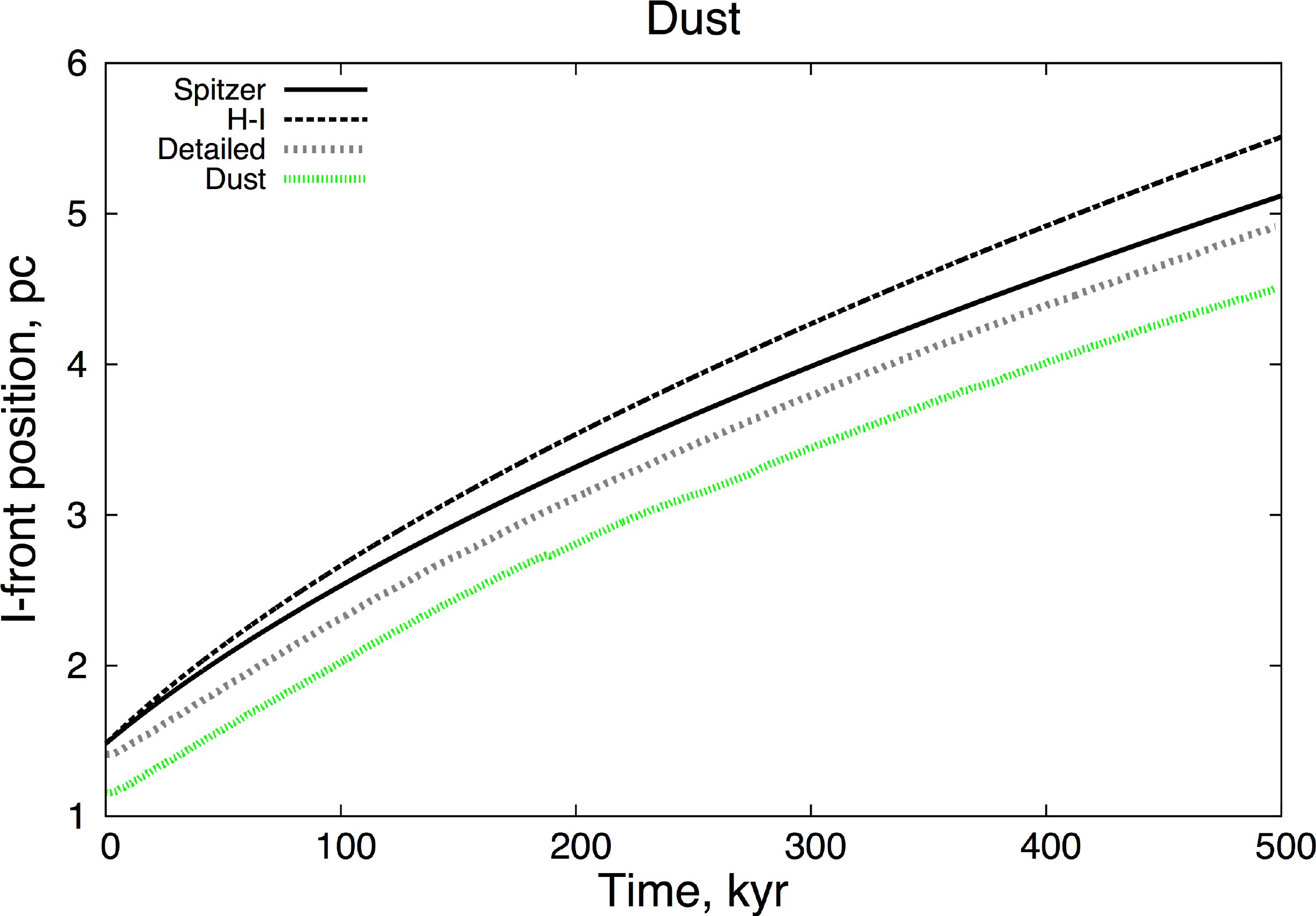}
	\includegraphics[width=9cm]{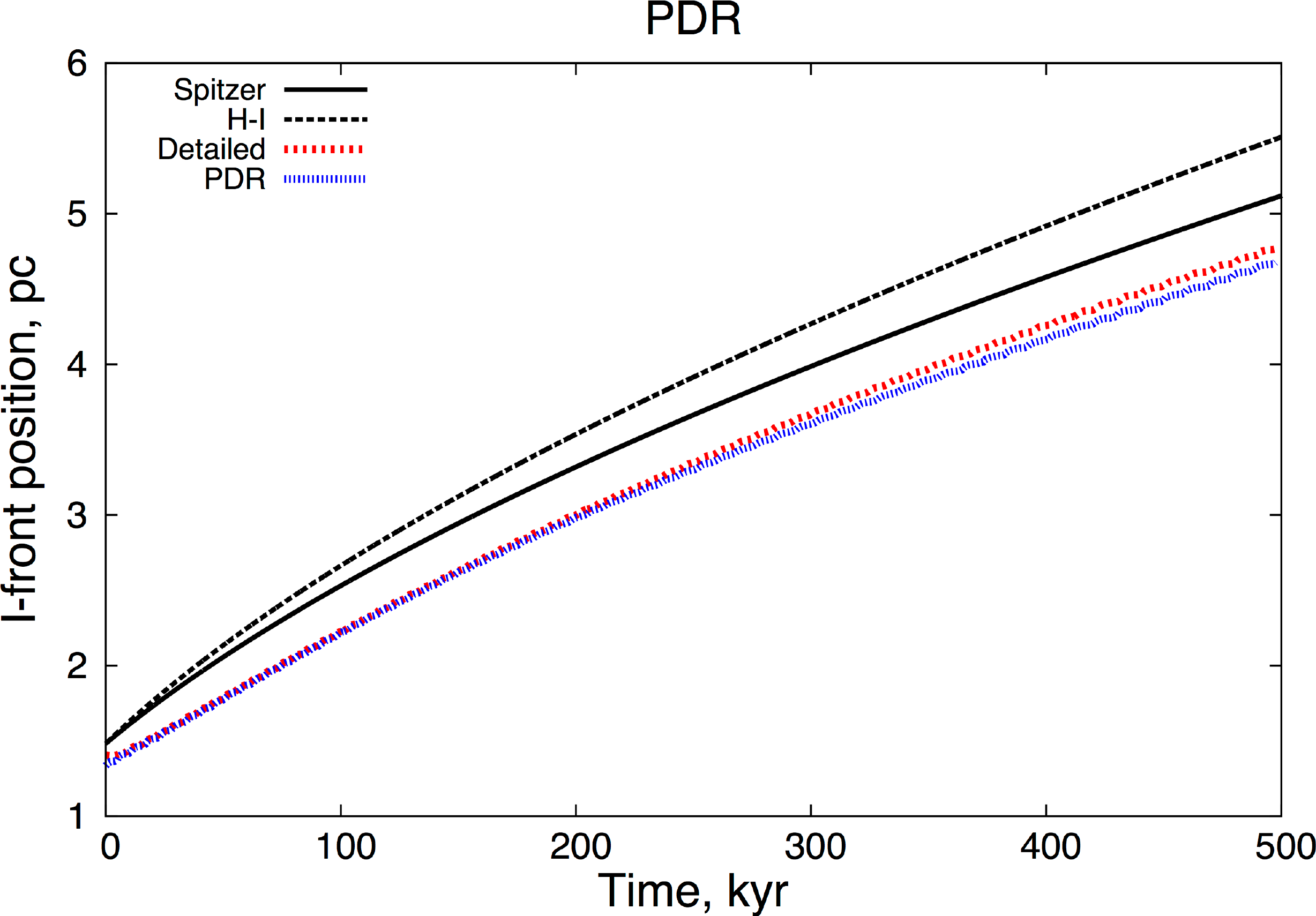}
	\caption{The ionisation front position as a function of time for all of the 500\,$m_{\rm{H}}$\,cm$^{-3}$ models in this paper. The top left panel compares the most simplified model with standard Monte Carlo photoionisation plus hydrodynamics. The top right panel includes detailed spectral models. Middle left varies the gas metallicity, middle right includes radiation pressure, bottom left includes dust and the bottom right includes PDR treatment. }
	\label{all2}
\end{figure*}

We compare the models by tracking the ionisation front position (defined as the point at which the hydrogen ionisation fraction is a half) as a function of time, this is shown for all low  and high density models in Figures \ref{all} and \ref{all2}. We now discuss each set of approximations/physical processes in turn.

\subsection{Simple versus full photoionisation}
\label{simple}
The simple photoionisation scheme assumes an ionised gas temperature of $10^4$\,K. If the temperature resulting from the  detailed photoionisation differs from this then the expansion rate of the H\,\textsc{ii} region will differ,  since the Str\"{o}mgren radius varies as $r_s\propto T^{0.27}$ (for a case B recombination coefficient) and at later times during the expansion varies as $r_{\textrm{I}}\propto T^{2/7}$ (see equation \ref{spitzer}). The simple scheme also considers hydrogen only gas, which means that the electron density compared to a calculation with helium and metals may also differ. Finally the simplified scheme also uses the OTS approximation and so there may be differences between the case-B recombination coefficient used (see equation \ref{alphaB}) and the net effective recombination coefficient for the calculation with multiple species.

The top left panels of Figures \ref{all} and \ref{all2} show the ionisation front position as a function of time for the Spitzer analytic solution (equation \ref{spitzer}) as well as for our simplified and full photoionisation calculations with \textsc{torus} for the 100 and 500\,m$_{\rm{H}}$\,cm$^{-3}$ density calculations. We also include the analytic expression describing the expansion given by  \cite{2006ApJ...646..240H} (labelled H--I),  which is derived by solving the equation of motion of the shell of material swept up by the expanding H\,\textsc{ii} region
\begin{equation}
	\frac{d(M\dot{R})}{dt} = 4\pi R^2\left(P_i - P_o\right)	
\end{equation}
which, under the assumption that $P_o \ll P_i$, results in  
\begin{equation}
	r_I = r_s\left(1 + \frac{7\sqrt{4}\, c_I t}{4\sqrt{3}\, r_s}\right)^{4/7}.
	\label{hi}
\end{equation}
The difference between the H--I and Spitzer solutions is that the former solves the equation of motion of the shell (thin shell approximation) and the latter considers pressure balance between the H\,\textsc{ii} region and ambient medium at each point in time). 

\subsubsection{Initial Str\"{o}mgren radii}
\label{stromsec}
Looking at the top left panels of Figures \ref{all} and \ref{all2}, the simplified models agree with the Str\"{o}mgren solution for the position of the onset of D--type expansion in both density regimes, however for the detailed model this radius is smaller by 7.2 and 6.5 per cent in the 100 and 500\,m$_{\rm{H}}$\,cm$^{-3}$ models respectively. Given that the number of ionising photons is the same in each model, the difference can only come from a difference in the electron density or recombination rate from equation \ref{stromgren}.

We compared the electron density for the two models at the onset of D--type expansion and found they are very consistent. The difference must therefore be arising from a higher effective recombination rate across all species relative to the case B recombination coefficient used to calculate the Str\"{o}mgren radius analytically. The recombination rate (equation \ref{alphaB}) is temperature dependent, so the difference could be explained if the simple and detailed photoionisation schemes result in different gas temperatures. We plot the gas temperature as a function of radius from the ionising source in Figure  \ref{simpfulltemp}. The simplified calculation results in higher average gas temperatures, implying a lower recombination rate and hence a larger Str\"{o}mgren radius (as  we observe in our models).  

In Figure \ref{photodet_explained}, which we will discuss more below, we compare our simplified and detailed photoionisation results against analytic solutions using gas temperatures representative of the model average in the ionised gas. When we do this the initial I-front position of the detailed photoionisation models is in good agreement with the Str\"{o}mgren equation. So we can conclude that the difference in Str\"{o}mgren radius is explained by a difference in the gas temperature and hence recombination rate.

We note that the H\textsc{ii} region radius calculated by \textsc{torus} for detailed photoionisation models  has been shown to agree with other photoionisation codes, e.g. \cite{2012MNRAS.420..562H} where we compare with the \textsc{cloudy} code \citep{1998PASP..110..761F}.


\subsubsection{Expansion rates}
\label{ratesimple}
We remind the reader that the top left panels of Figures \ref{all} and \ref{all2} show the ionisation front position as a function of time of the models being discussed presently. The StarBench\footnote{{https://www.astro.uni-bonn.de/sb-ii/}}  code comparison project D--type expansion test (which uses the simplified photoionization physics and towards which \textsc{torus} contributed results) found that numerical codes agree with the Spitzer solution early on, but eventually overtake it to agree more closely with the Hosokawa-Inutsuka solution. This behaviour is reproduced in the simple photoionization models in this paper, though the expansion is slightly slower than expected in the lower density regime.  We expect that this is due to the comparatively low resolution (256 cells) used for the simulations in this paper compared to the 1024 cells used in the 1D simulations of \cite{2015arXiv150705621B}. Although we could increase the resolution for the simplified calculations (indeed our contributions to the StarBench D-type test were at higher resolution) this would make consistent resolution calculations including the PDR very computationally expensive. 

\begin{figure}
	\includegraphics[width=9cm]{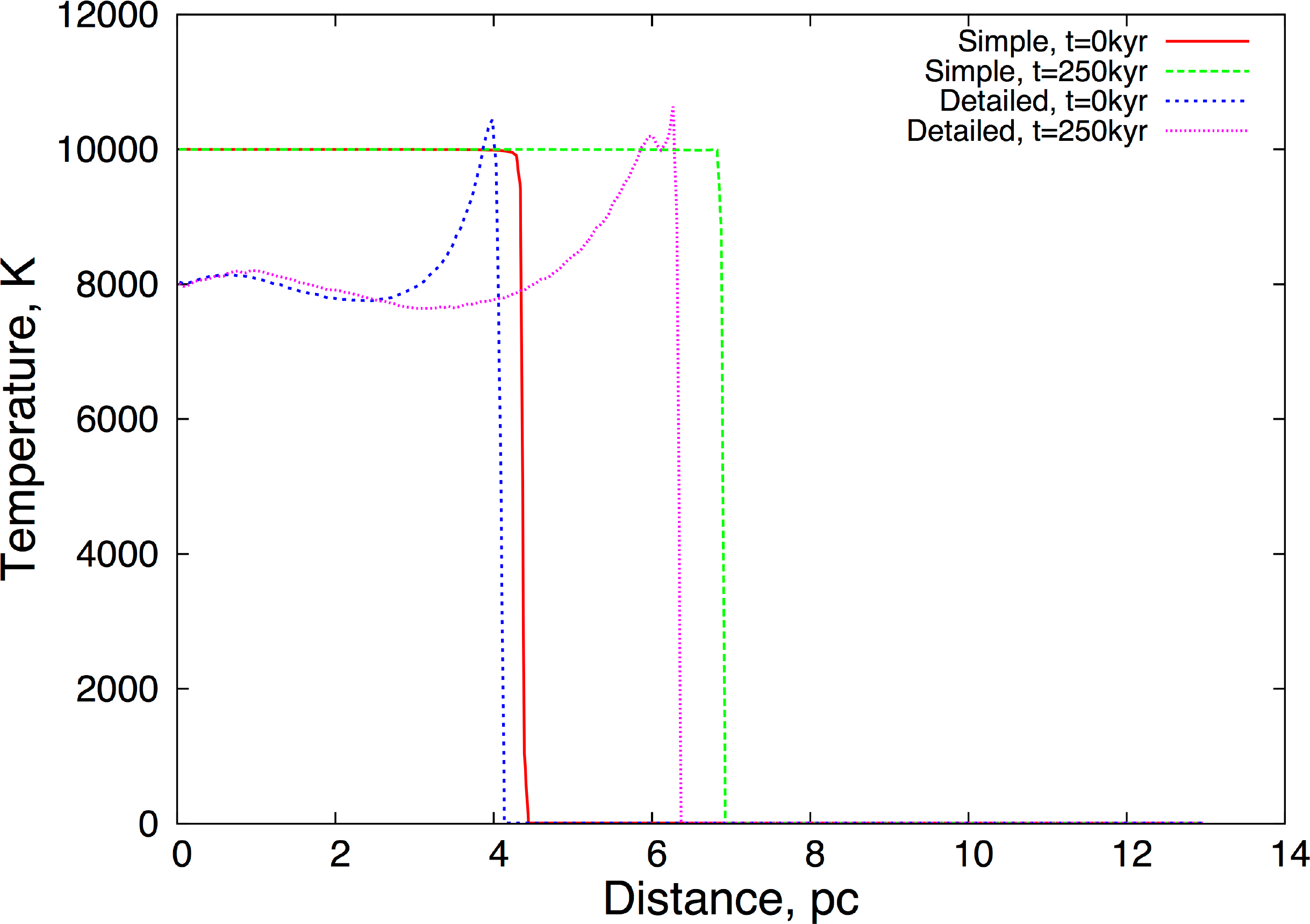}	

	\includegraphics[width=9cm]{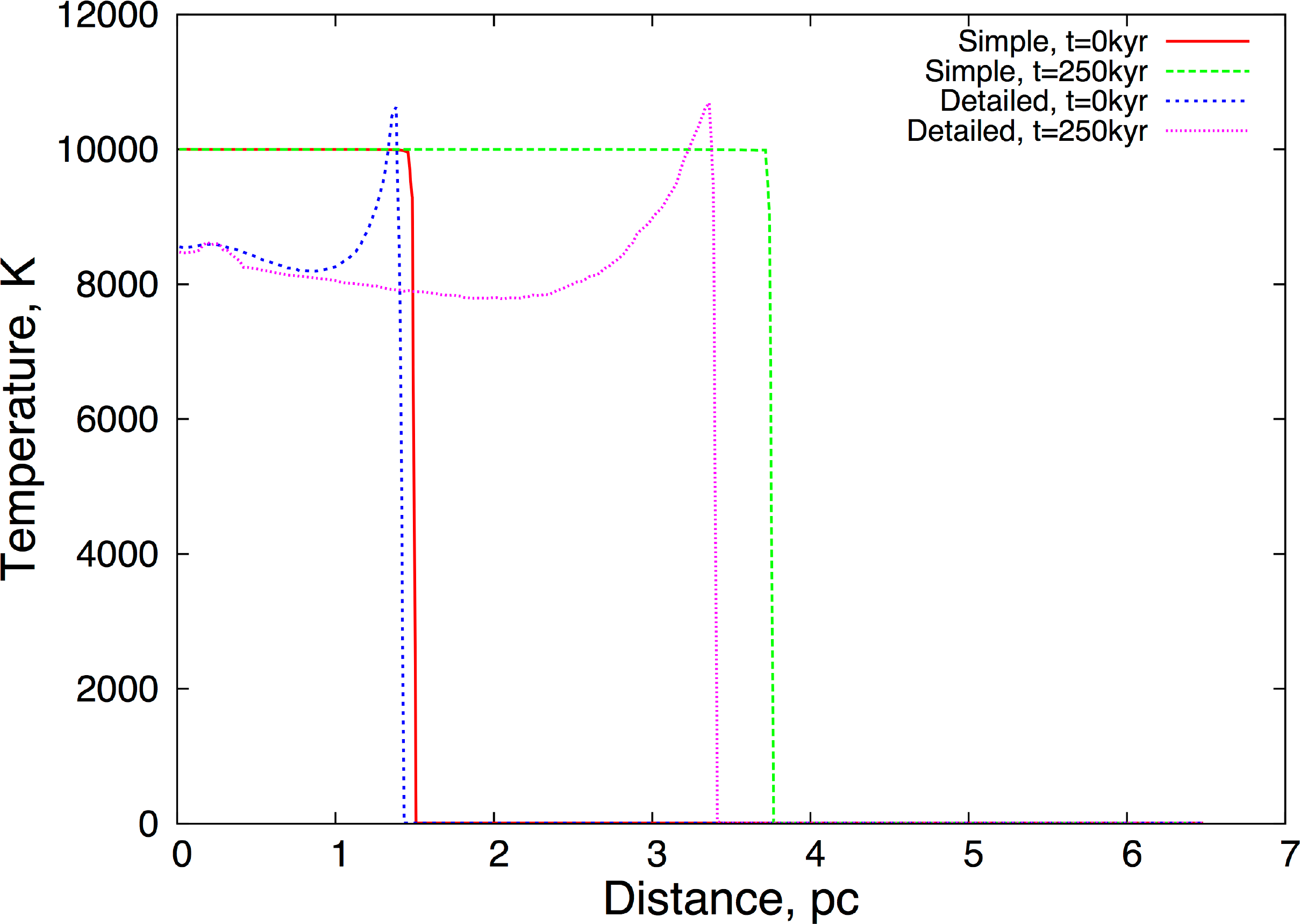}
	\caption{The radial temperature structure of the simple and detailed photoionisation models at the simulation start time and after 250\,kyr. The upper and lower panels are for the 100 and 500\,m$_{\textrm{H}}$\,cm$^{-3}$ ambient density models respectively.}
	\label{simpfulltemp}
\end{figure}

\begin{figure}
	\includegraphics[width=9cm]{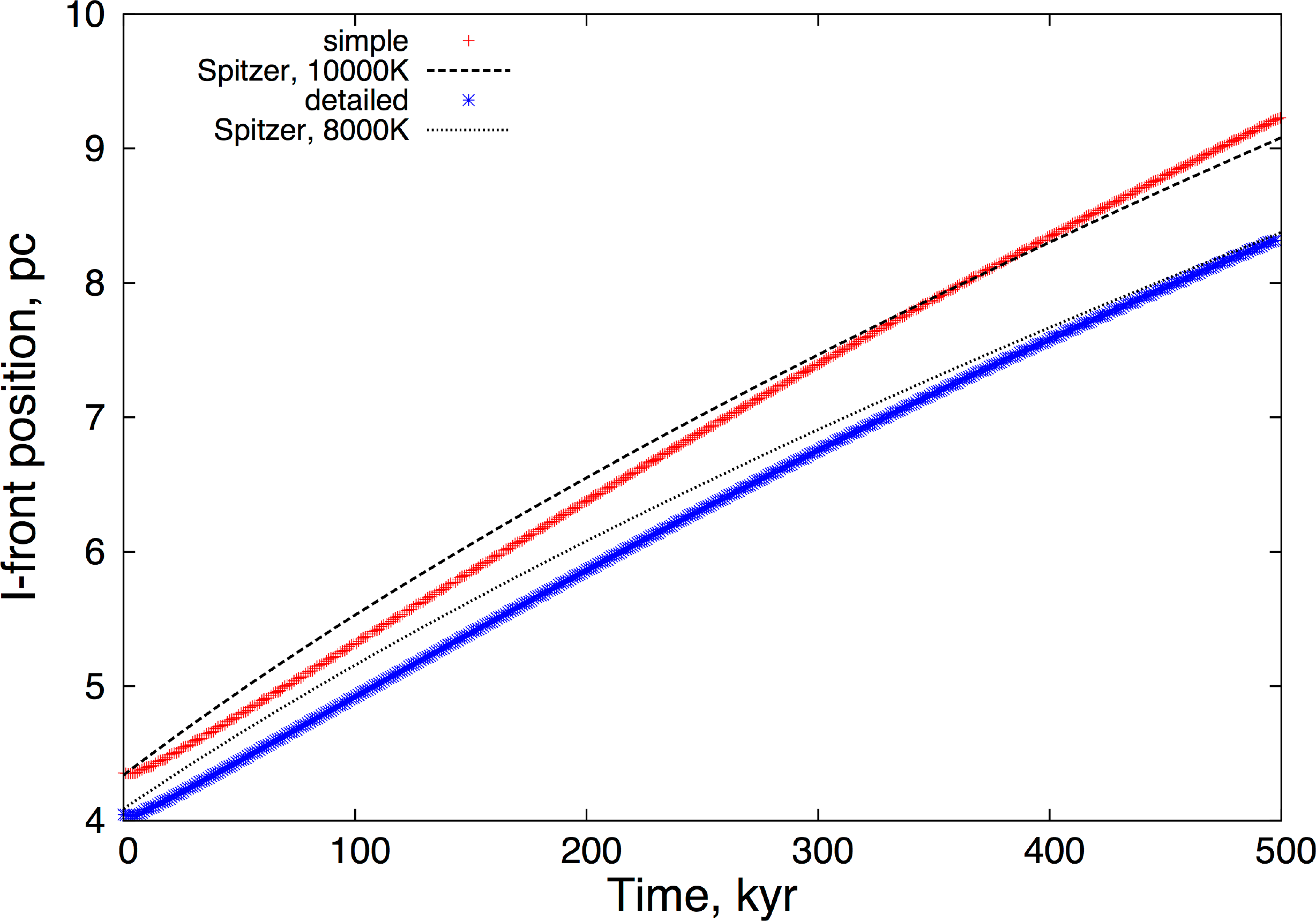}
	\includegraphics[width=9cm]{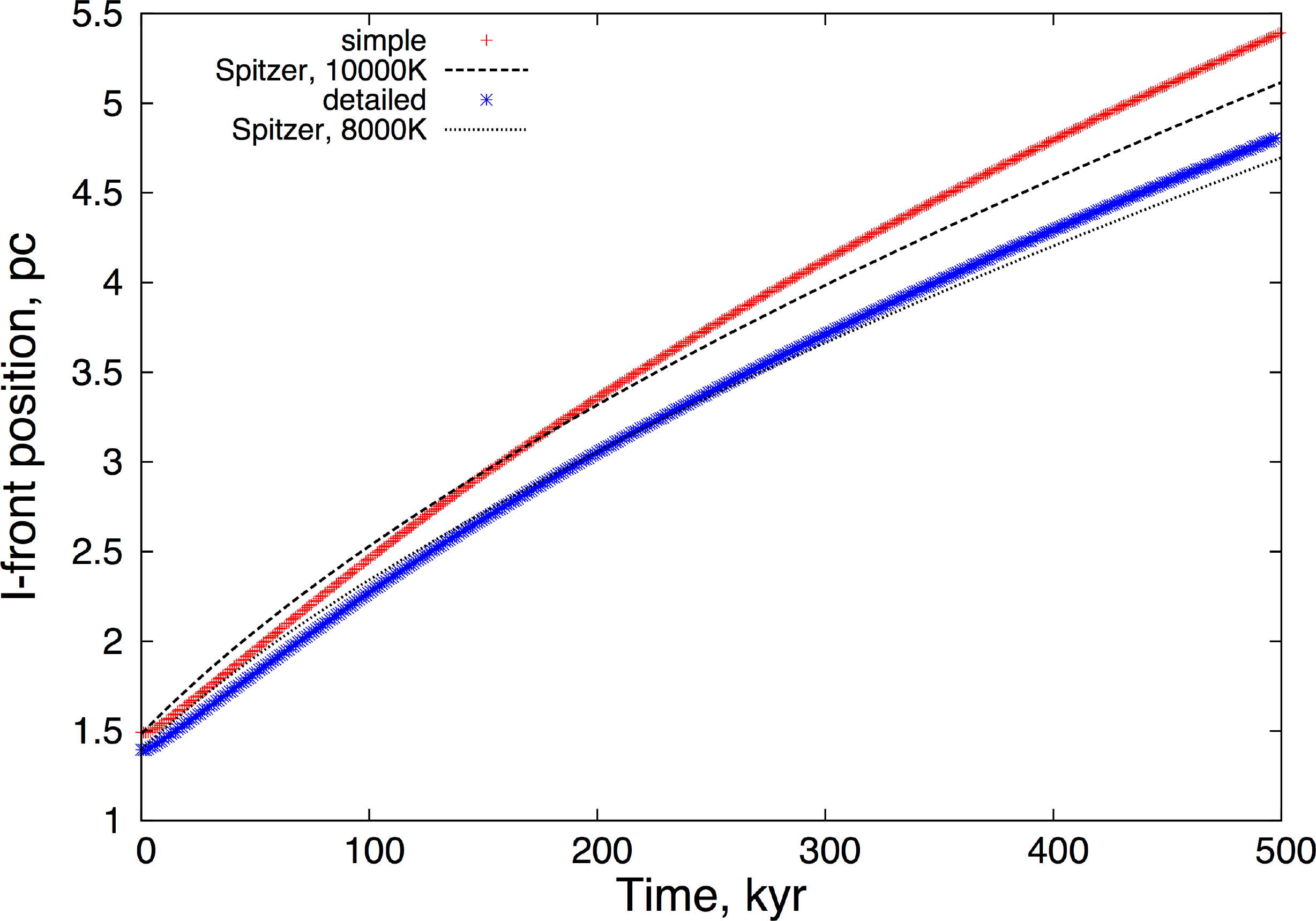}	
	\caption{Demonstrating that the differences between the simplified and detailed photoionisation models can be explained by their different ionised gas temperatures. The upper and lower panels show results for the 100 and 500\,m$_{\textrm{H}}$\,cm$^{-3}$ density models respectively. The analytic solutions are just the Spitzer equation with an ionised gas temperature of 10000\,K and 8000\,K for the simple and detailed photoionisation models respectively. }
	\label{photodet_explained}
\end{figure}

The effect of moving to full photoionisation is not only to reduce the radius of D-type onset (see section \ref{stromsec}), but also to reduce the expansion \textit{rate} of the H\,\textsc{ii} region. In Figure \ref{simpfulltemp} we show the temperature of the detailed and simple models just before the onset of D-type expansion and after 250\,kyr. Throughout most of the H\,\textsc{ii} region the temperature in the ionised gas of the full photoionisation model is around 2000\,K lower than that of the simplified model, except for the temperature peak near to the ionisation front. These cooler average temperatures are due to the efficient forbidden line cooling, which is only included in the detailed model. The peak in temperature close to the ionisation front arises where coolants with higher ionisation potentials recombine. The simple and detailed models have average gas temperature of $10^{4}$ and 8000\,K  respectively. In Figure  \ref{photodet_explained} we use these average gas temperatures in the Spitzer solution and compare with our numerical results. This demonstrates that differences in the D-type expansion between the simple and detailed photoionisation models are simply explained by the differences in the ionised gas temperature.



\begin{table}
 \centering
 \begin{minipage}{180mm}
  \caption{Ionizing fluxes for the different stellar spectral models.}
  \label{stars}
  \begin{tabular}{@{}l c c@{}}
  \hline
   Stellar Model & Metallicity & ionising photons ($\times10^{49}$/s)\\
  \hline   
   Blackbody & -  & 2.69\\  
   \textsc{tlusty} & 0.5 & 2.60\\
   \textsc{tlusty} & 1 & 2.63 \\
   \textsc{tlusty} & 2 & 2.70\\
\hline
\end{tabular}
\end{minipage}
\end{table}

\subsection{Stellar spectral models}
Although the different model spectra shown in Figure \ref{BBvsSpec} appear significantly different, the integrated ionising flux over the spectrum is actually very similar in each case for stars of this effective temperature \citep[see table \ref{stars} in this paper and Figure 15 from][]{2003ApJS..146..417L}. The expansion of H\,\textsc{ii} regions about stars with these different spectra is therefore very similar, as we show in the top right panels of Figures \ref{all} and \ref{all2}. The difference in the initial Str\"{o}mgren radius is negligible and after 500\,kyr of expansion the difference in H\,\textsc{ii} region radius is only of order 1--3 per cent in both density regimes.

The lower panel of Figure \ref{BBvsSpec}  shows that harder photons are produced in the \textsc{tlusty} models, however this has very little effect on the ionisation state or dynamical evolution, only slightly broadening the ionisation front. This is in agreement with what was found by 
\cite{2012MNRAS.420..562H} in 3D models of radiatively driven implosion, where the effect of including harder radiation was isolated. Although hard radiation can have important consequences at galactic or cosmological scales \citep[e.g.][]{2004MNRAS.348..753S, 2006MNRAS.368.1885I}, on scales of up to tens of parsecs we find that its effect is negligible. 

\subsection{Gas metallicity}
The gas metallicity determines the amount of metal line cooling and therefore the temperature in the ionised gas. Since the Str\"{o}mgren radius varies as T$^{0.27}$ and at later times  $r_I\propto T^{2/7}$ the expansion rate of H\,\textsc{ii} regions is expected to be metallicity dependent.  We find that this is the case in our models, which we show in the middle left panels of Figures \ref{all} and \ref{all2}. Included are calculations at the base metallicity (which we call $z_G = 1$, that is the same as that used in the photoionisation HII40 Lexington benchmark) as well as $z_G = 0.5$ (representative of earlier universe, LMC) and $z_G = 2$ (representative of the Galactic centre). The lower metallicity expansion is much faster than the higher metallicity, giving a $z_G = 0.5$, 500\,kyr H\,\textsc{ii} region radius larger than that for  $z_G = 2$ by 18 and 16 per cent in the 100 and 500\,m$_H$\,cm$^{-3}$ models respectively.

\begin{figure}
	\includegraphics[width=8cm]{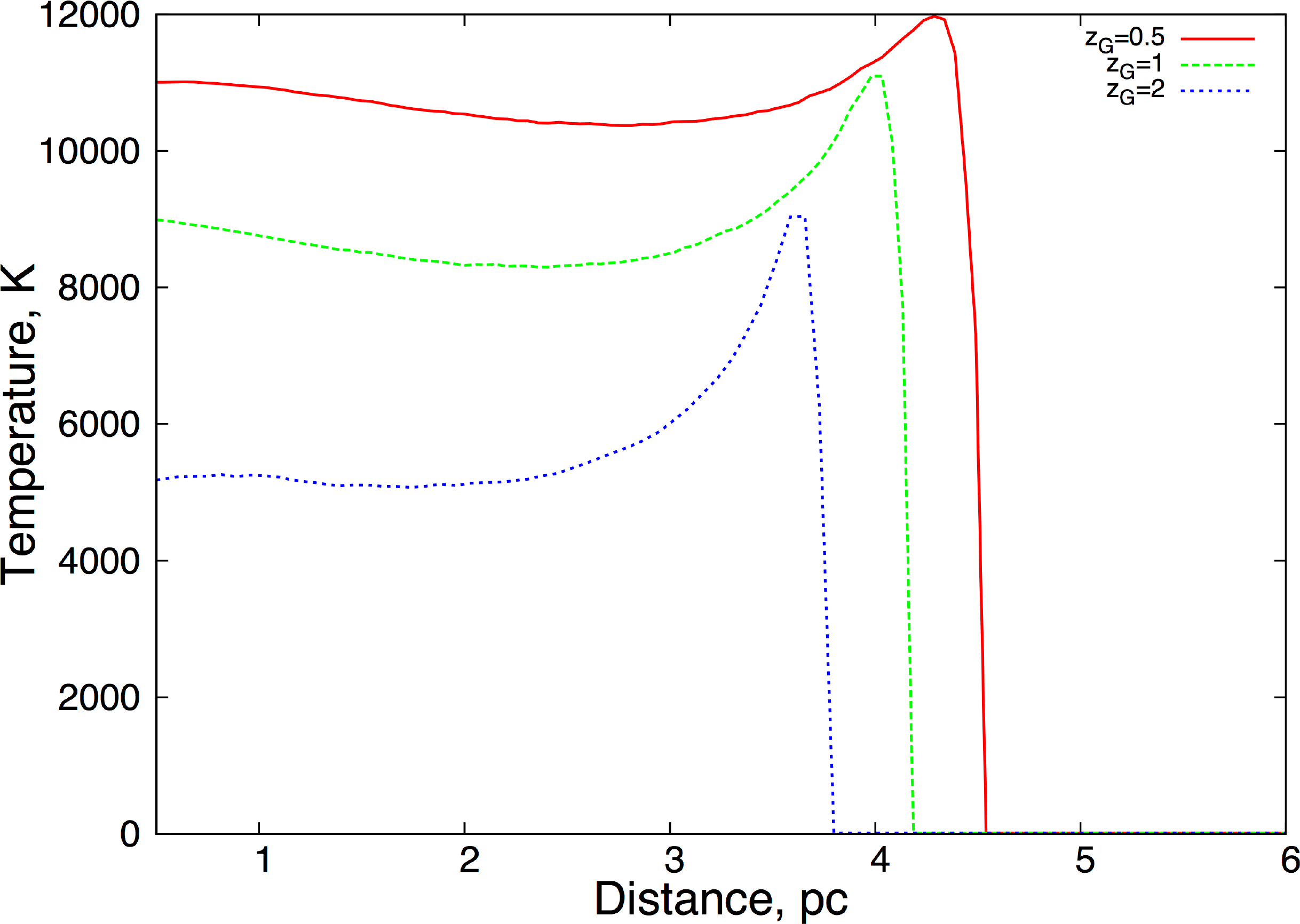}
	\caption{The gas temperatures in the 100\,m$_{\textrm{H}}$\,cm$^{-3}$ model at the onset of D-type expansion for gasses of different metallicity: 0.5 (red), 1 (green) and 2 (blue) times the HII40 Lexington benchmark metallicity. }
	\label{tempmetfix}
\end{figure}

\begin{figure}
	\includegraphics[width=8cm]{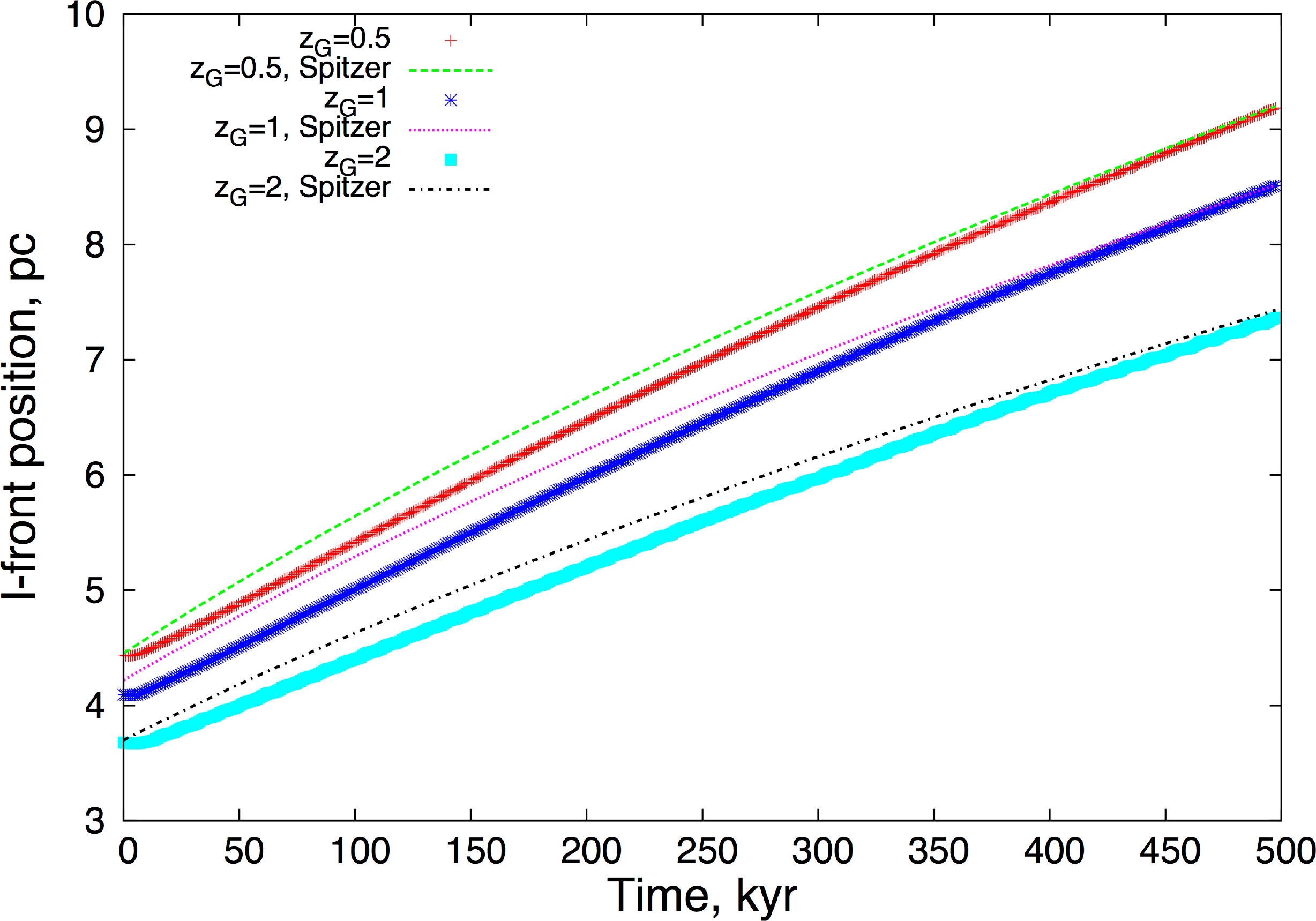}
	\includegraphics[width=8cm]{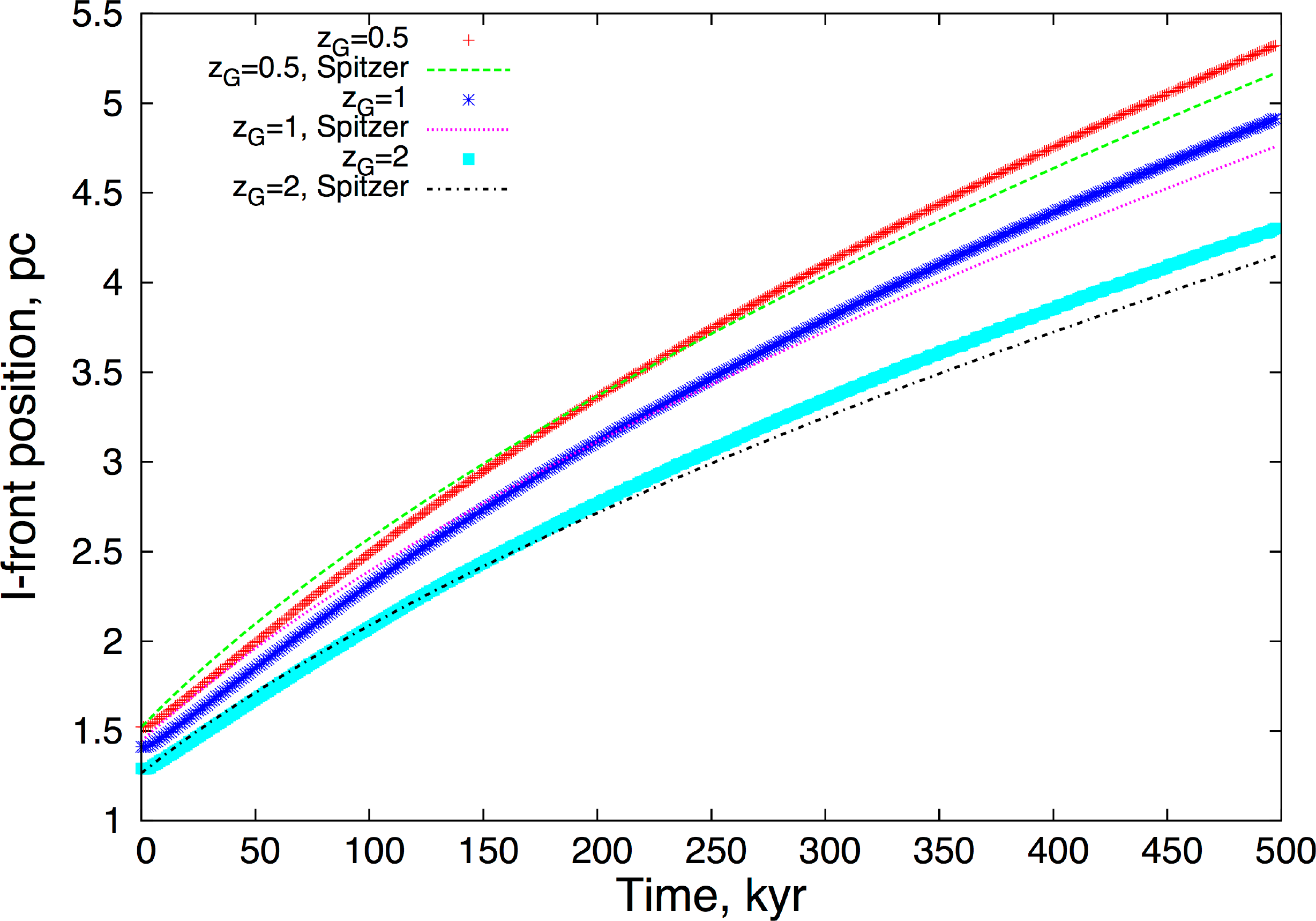}
	\caption{The numerical simulations compared with analytical results with fitted gas temperatures. The upper and lower panels are for the 100 and 500\,m$_{\textrm{H}}$\,cm$^{-3}$ models respectively.}
	\label{gasmet_fits}
\end{figure}

In  Figure \ref{tempmetfix} we show the radial temperature profile of the different metallicity H\,\textsc{ii} regions at a time close to the onset of D--type expansion. The temperature in the ionised gas and the Str\"{o}mgren radius are both decreasing  as a function of gas metallicity.  This is because at lower metallicity there is weaker metal line cooling. Figure  \ref{gasmet_fits} shows the ionisation front position as a function of time for the models of different gas metallicity, as well as analytic plots which are generated from the Spitzer solution and Str\"{o}mgren radius, only setting the ionised gas temperature to the average seen in Figure \ref{tempmetfix}. The agreement is good enough for us to conclude that the differences in H\,\textsc{ii} region expansion rate with metallicity are dominated by differences in the ionised gas temperature. 

Since there is good agreement between the theoretical expressions and simulations when the correct temperature is used, we fit our temperature as a function of metallicity to obtain an approximate,  simplified thermal calculation that accounts for the gas metallicity, improving on the simple prescription given by equation \ref{quicktherm}
\begin{equation}
	T = T_n + \left[1.1\times10^{4} - 3.8\times10^3(z/z_o - 0.5)^{0.839} - T_n\right]\eta
	\label{mettemp}
\end{equation}
where $z/z_o$ is the gas metallicity relative to that of the Lexington benchmark (a standard Milky Way star forming region abundance), $T_n$ is the neutral gas temperature and $\eta$ is the hydrogen ionisation fraction. 
An example potential application of this approximate simplified temperature calculation would be to use it 3D models such as those of \cite{2013MNRAS.431.1062D} or \cite{2013MNRAS.435..917W} (both of which use equation \ref{quicktherm}) to study how radiative feedback affects molecular clouds in regions of different metallicity, or as a function of cosmic time. \\

In reality the gas and stellar metallicities will not be independent. We  investigate them separately in this paper to identify which of the two components dominates any differences in the H\,\textsc{ii} region evolution. Based on our simulations we find that feedback dynamics are much more sensitive to the gas metallicty than changes in the stellar metallicity.


\subsection{Radiation pressure}
The ionisation front evolution for models that include radiation pressure is given in the middle right panels of Figures \ref{all} and \ref{all2}. The radiation pressure simulations are just the ``detailed'' models with the addition of the radiation pressure formulation developed by \cite{2015MNRAS.448.3156H}. In both density regimes the effect of including radiation pressure is essentially zero ($<$1 per cent).  This is in agreement with \cite{2014MNRAS.439.2990S}, who find that radiation pressure effects are secondary to photoionisation in D--type expansion models. Further note that \cite{Raga2015} finds that hundreds of O stars are required for radiation pressure to become important to the system dynamics, such as in the Tarantula nebula in the LMC.  

\subsection{Dust}
Dust absorption reduces the ionising photon budget and hence the size of the Str\"{o}mgren radius (see equation \ref{stromgren}). In Figure \ref{dusttest} we also saw that it increases the temperature downstream of the ionisation front in the \textsc{mocassin} results (though this is missing from our treatment of the dust in this paper, the consequences of this downstream heating effect will be gauged in our discussion of PDR heating in section \ref{pdrsec}). The ionisation front expansion for the dust models is given in the bottom left panels of Figures \ref{all} and \ref{all2}. As we found when testing in Figure \ref{dusttest}, inclusion of dust leads to the initial Str\"{o}mgren radius being reduced. This affects the expansion rate of the H\,\textsc{ii} region at times when the Str\"{o}mgren term is important, i.e. when $\frac{7\, c_{\textrm{I}} t}{4\, r_{\textrm{s}}} \lesssim 1$. Although the ionising photon budget is reduced, the gas temperature remains very similar (see Figure \ref{dusttest}) and so at later times when $r_I\propto T^{2/7}$ the expansion behaves in a similar manner either with or without dust. The effect of dust, minus any effect from downstream heating, is therefore simply to reduce the radius at which D--type expansion begins, resulting in a smaller H\,\textsc{ii} compared to models without dust.

\subsection{Photodissociation regions}
\label{pdrsec}

Models including treatment of PDRs are the most computationally intensive in this paper. With these models we can follow ionised hydrogen, through to the atomic and molecular transitions (and can do the same for, e.g. carbon and CO and many other species). The PDR does not affect the temperature in the ionised gas, but leads to heating of up to around a few hundred Kelvin in the neutral gas out to relatively large distances (potentially many parsecs). 

In the derivation of Spitzer and H-I equations, there is a term involving the difference in the pressures in the ionised and neutral gas $P_i - P_o$, where $P_o$ is dropped assuming $P_o << P_i$. If the PDR heating raises the temperature enough then this assumption will no longer be appropriate and the expansion will be slower. In \cite{2015arXiv150705621B} the full form of the Spitzer and Hosokawa-Inutsuka expressions are presented without neglecting the external pressure, 
\begin{equation}
	\frac{1}{c_i}\frac{d r_I }{dt} = \left(\frac{r_s}{r_I}\right)^{3/4} - \frac{\mu_iT_o}{\mu_oT_i}\left(\frac{r_s}{r_I}\right)^{-3/4} 
	\label{fullSpitzer}
\end{equation}
and 
\begin{equation}
\frac{1}{c_i}\frac{dr_I}{dt} = \sqrt{\frac{4r_s^{3/2}}{3r_I^{3/2}} - \frac{\mu_iT_o}{\mu_oT_i}}
	\label{fullHI}
\end{equation}
respectively. These expressions need to be evaluated numerically. In each case it is the $\mu_iT_o/\mu_oT_i$ term (which is generally expected to be small) that is dropped to arrive at the Spitzer or Hosokawa-Inutsuka results. PDR (or dust) heating may change this though.

The evolution of the H\,\textsc{ii} region radius for these PDRRHD models is given in the bottom right panels of Figures \ref{all} and \ref{all2}. In both density regimes the effect of the PDR is to marginally slow the H\,\textsc{ii} region expansion. The PDR heated gas close to the ionisation front is $\sim300$\,K in both density regimes. Substituting this temperature and an ionised gas temperature of 8200\,K into equation \ref{fullSpitzer} and comparing with equation \ref{spitzer} gives an expected difference at 500\,kyr of about 1 per cent in the H\,\textsc{ii} region extent. This is in agreement with the difference found in our simulations. Given the small difference from downstream heating in the PDR model, we also expect the similar heating from dust (see Figure \ref{dusttest}) to have a negligible effect.

\subsection{Consequences for RHD modelling}
We have considered the effect of many different microphysical processes that are not normally included in RHD simulations because they make the calculation much more expensive (for the models including PDR treatment, by an order of magnitude or more). Studying all of these processes at once in a systematic way has the added value that they will not be studied across multiple papers and we can immediately see which processes are important and in what ways current simplified models are deficient.

\subsubsection{The state of simplified models}
Compared to the simplified models in this paper, every other process (except lowering the gas metallicity) results in a smaller H\,\textsc{ii} region. It is therefore possible that RHD simulations of more complex geometrical systems (such as turbulent star forming regions) using simplified microphysics are overestimating the power of the radiation field. In the context of, for example, radiative feedback and triggered star formation, this could mean that the ability of radiative feedback to compress clouds and trigger star formation might be reduced. Conversely its ability to disperse clouds, halting star formation is also reduced. 

The metallicity is also particularly important, with approximately a 20 per cent difference in H\,\textsc{ii} region radius after 500\.kyr between models with metallicities 0.5 and 2 times our base metallicity. Radiative feedback may therefore play a substantially different role in the earlier universe or LMC compared to somewhere like the Galactic centre. Equation \ref{mettemp}, a simple thermal prescription which accounts for the gas metallicity, could be used to investigate this in 3D models.

\subsubsection{Hierarchy of processes when modelling D-type expansion of galactic H\,\textsc{ii} regions}
\begin{figure}
	\includegraphics[width=8.5cm]{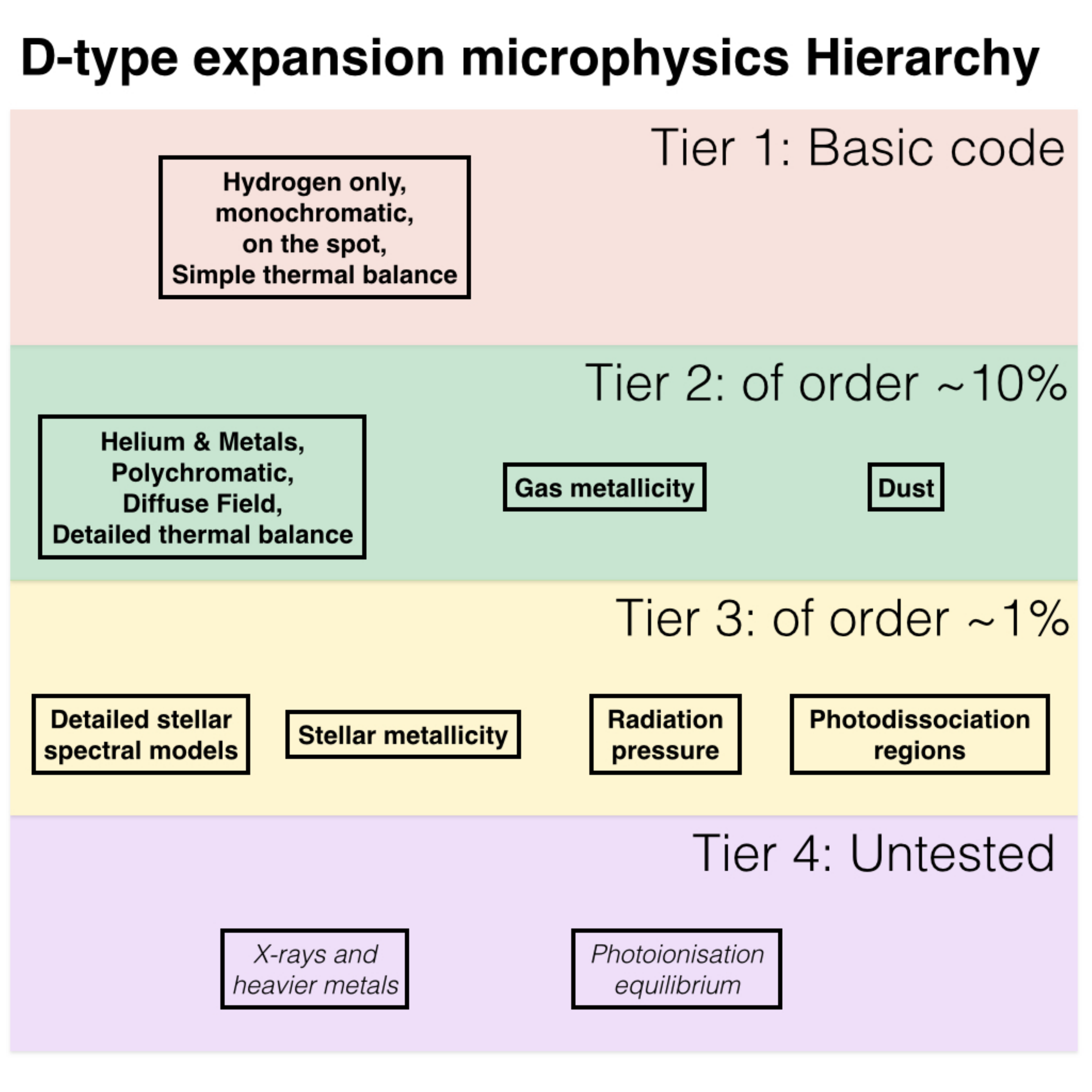}
	\caption{A hierarchy of importance of physical processes \textit{for the D--type expansion model considered in this paper}. There will be other scenarios where some of these components become significantly more important, e.g. massive star clusters such as the tarantula nebula in the LMC where radiation pressure plays an important role. Note that the percentage difference quoted is only after 500\,kyr of H\,\textsc{ii} region evolution. }
	\label{hierarchy}
\end{figure}

We have shown here that if only the dynamical evolution of the system needs to be modelled (as opposed to chemistry or synthetic observations) there is not necessarily much value in spending time implementing, for example, detailed spectral models or PDR treatment, when the effect on the dynamical evolution of the system is negligible. We have therefore constructed a simple hierarchy of processes that we find to be important for modelling radiative feedback in star forming regions, which we show in Figure \ref{hierarchy}. Of course the effects will not be negligible in every scenario (for example radiation pressure is important early on in the formation of massive stars). 

The hierarchy consists of 4 tiers. Tier 1 is the standard basic RHD (photoionisation + hydrodynamics) code. Tier 2 includes processes that alter the H\,\textsc{ii} region extent at 500\,kyr by around 10 per cent or greater. Tier 2 includes processes that alter the H\,\textsc{ii} region extent at 500\,kyr by of around 1 per cent and tier 4 processes are untested. If improving the microphysics in a photoionisation + hydrodynamics code for non-cosmological applications this should offer a useful reference as to which processes are most important to include. We note that in reality the gas  and stellar metallicities are linked. Our hierarchy does not suggest that they are decoupled, rather that if the overall metallicity varies it is the changes in the gas properties rather than the stellar spectrum that affect changes in the dynamical evolution of the system.

\subsection{Untested features}
There are some processes that we identify that might also affect the D-type expansion of galactic H\,\textsc{ii} regions that we do not treat in this paper. 

\subsubsection{X-rays and heavier metals}
Treatment of X--ray photons and a wider range of metal species might affect the ionisation and temperature structure in the H\,\textsc{ii} region. Such treatment has been implemented in the Monte Carlo photoionisation code \textsc{moccasin} \citep{2008ApJS..175..534E} but is not currently available in \textsc{torus}. Inclusion of X--rays may lead to increased temperatures in the H\,\textsc{ii} region since atoms/ions with higher ionisation potentials will be ionised. This would lead to 
an increase in the expansion rate as we found with the lower gas metallicity models.

\subsubsection{Non-equilibrium photoionisation}
Including non--equilibrium photoionisaiton is not expected to modify the results of the calculations in this paper. It would allow us to capture the R--type expansion of the ionised gas that occurs prior to the D--type expansion that we study here, however R--type expansion occurs rapidly and with little disruption to the density field. Although we do not expect this assumption to affect the resulting density field  the results of  \cite{2012A&A...546A..33T} show that it will be important to test the effects of non-equilibrium photoionisation on shadowed regions.\\

\section{Summary and conclusions}
We have used the Monte Carlo radiation transport and hydrodynamics code \textsc{torus} and its PDR counterpart \textsc{torus-3dpdr} to study the effects of different microphysics on the D--type expansion of H\,\textsc{ii} regions. We have compared analytic solutions with the simplest RHD models, calculations including multiple species, the diffuse field, polychromatic radiation, detailed spectral models, different gas metallicity, radiation pressure, dust and PDR's.  Each of these processes has been treated directly rather than using a simple parameterisation. We draw the following main conclusions from this work: \\

\noindent 1) The critical factor that affects the H\,\textsc{ii} region dynamics when different microphysics is treated is the gas temperature, which affects the Str\"{o}mgren radius as $r_s\propto T^{0.27}$ and the later time expansion rate as $r_I\propto T^{2/7}$. The  Str\"{o}mgren radius directly affects the expansion rate at times up until $\frac{7\, c_{\textrm{I}} t}{4\, r_{\textrm{s}}} >> 1$. \\

\noindent2) Compared to simplified (hydrogen only, monochromatic, on the spot approximation) models, cooler average gas temperatures resulting from full Monte Carlo photoionisation can change the extent of  the H\,\textsc{ii} region after 500\,kyr by about 10 per cent (see conclusion 1). This implies that simplified RHD models are overestimating the power of radiative feedback, be it to induce star formation or disperse gas in star forming regions. \\

\noindent 3) Varying the gas metallicity changes the gas amount of metal line cooling and hence the gas temperature. Changing the metallicity by a factor of 4 can also lead to a 10--20 per cent difference in the H\,\textsc{ii} region extent after 500\,kyr. The Spitzer D--type expansion solution matches our metallicity dependent models well if it uses the average gas temperatures from our simulation. We fit our models to produce a simplified thermal prescription that can be used in 3D models to incorporate heating effects of multi-species gas at varying metallicity. This is at negligible additional computational cost. It will also allow us to test how deficient simplified schemes are in more complex 3D applications.  \\

\noindent 3) Dust absorption reduces the ionising photon budget, meaning that  the onset of D-type expansion occurs at smaller radii (since $r_s \propto N_{\rm{ly}}^{1/3}$). The H\,\textsc{ii} region temperature is similar, so the later time expansion rate (when the Str\"{o}mgren radius term becomes less important) is similar with or without dust.\\

\noindent 4) Using detailed stellar spectral models does not affect the simulation much (of order 1 per cent difference in the ionised gas extent at 500\,kyr), since the ionising flux remains approximately constant and the effect of any harder radiation is just to smear out the ionisation front. Radiation pressure is also found to be dominated by the effects of photoionisation, in agreement with previous studies such as \cite{2014MNRAS.439.2990S}. PDR heating of the downstream medium only marginally affects the H\,\textsc{ii} region expansion (by $\sim$\,1 per cent at 500\,kyr), in agreement with the difference expected from the analytic equations.\\

\noindent 6) We develop a hierarchy of processes for modelling the D-type expansion of H\,\textsc{ii} regions, where additional physics beyond the most simple model is graded based upon its effect upon the expansion. This offers guidance for where to focus in future development of numerical models. \\


\section*{Acknowledgments}
TJHaw is funded by the STFC consolidated grant ST/K000985/1. TJHarries acknowledges funding via an STFC consolidated grant ST/J001627/1. The work of TGB was funded by STFC grant ST/J001511/1.

The calculations presented here were performed using the Darwin Data Analytic system at the University of Cambridge, operated by the University of Cambridge High Performance Computing Service on behalf of the STFC DiRAC HPC Facility (www.dirac.ac.uk). This equipment was funded by a BIS National E-infrastructure capital grant (ST/K001590/1), STFC capital grants ST/H008861/1 and ST/H00887X/1, and DiRAC Operations grant ST/K00333X/1. DiRAC is part of the National E-Infrastructure. This work was also supported by initial exploratory calculations performed using the DiRAC HPC facility machines Zen (Exeter) and Complexity (Leicester). 

We would like to acknowledge the Nordita program on Photo-Evaporation in Astrophysical
Systems (June 2013) where part of the work for this paper was carried
out. This work also benefitted from discussions held at the Starbench (Exeter, 2013) and Starbench-2 (B\"{o}nn, 2014) code comparison project meetings. We also thank Barbara Ercolano for her assistance in comparing \textsc{torus} with \textsc{mocassin} in the dust decoupled HII40 Lexington benchmark.

\bibliographystyle{mn2e}
\bibliography{molecular}

\bsp

\label{lastpage}

\end{document}